\documentclass[amsmath,amssymb,11pt]{article}
\usepackage{jheppub2}
\pdfoutput=1

\usepackage{amsmath,amssymb,amsthm,mathrsfs,bbm,bm}
\usepackage{latexsym,amscd,amsbsy,amsfonts,dsfont}
\usepackage{multirow}
\usepackage{graphicx}
\usepackage{xcolor}
\usepackage{caption}
\usepackage{comment}

\newcommand*{\affmark}[1][*]{\textsuperscript{#1}} 

\makeatletter
\DeclareRobustCommand\widecheck[1]{{\mathpalette\@widecheck{#1}}}
\def\@widecheck#1#2{%
    \setbox\z@\hbox{\m@th$#1#2$}%
    \setbox\tw@\hbox{\m@th$#1%
       \widehat{%
          \vrule\@width\z@\@height\ht\z@
          \vrule\@height\z@\@width\wd\z@}$}%
    \dp\tw@-\ht\z@
    \@tempdima\ht\z@ \advance\@tempdima2\ht\tw@ \divide\@tempdima\thr@@
    \setbox\tw@\hbox{%
       \raise\@tempdima\hbox{\scalebox{1}[-1]{\lower\@tempdima\box
\tw@}}}%
    {\ooalign{\box\tw@ \cr \box\z@}}}
\makeatother

\numberwithin{equation}{section}  
\allowdisplaybreaks 

\title{Love at the String Scale: Tidal Deformability Across the Black Hole--String Transition}

\author{Roberto Emparan\affmark[1,2],}
\emailAdd{emparan@ub.edu}
\author{Luis Lehner\affmark[3],}
\emailAdd{llehner@perimeterinstitute.ca}
\author{Stefano Trezzi\affmark[2]}
\emailAdd{strezzi@icc.ub.edu}

\affiliation{
\affmark[1]Institució Catalana de Recerca i Estudis Avançats (ICREA),
 Passeig Lluís Companys, 23, 08010 Barcelona, Spain\\
\affmark[2]Departament de Física Quàntica i Astrofísica and
  Institut de Ciències del Cosmos,
 Universitat de Barcelona, 08028 Barcelona, Spain\\
 \affmark[3]Perimeter Institute for Theoretical Physics, 31 Caroline Street North, Waterloo, ON N2L 2Y5, Canada\\
}

\abstract{
Tidal deformability provides a sensitive probe of the structure of compact objects, and black holes are exceptional in having vanishing static Love numbers in four-dimensional Einstein gravity. We ask what happens to this fine-tuned rigidity across the black hole--string transition, where the same states are expected to admit a weakly coupled description as a self-gravitating highly excited string, or ``string star''. We compute the static tidal Love numbers of the Horowitz--Polchinski (HP) string star for multipoles $\ell=2,3,4$ in $D=4,5,6$. The response is non-zero in all cases, as in the available $\alpha'$-corrected black hole results, so the zero-Love structure of four-dimensional Einstein gravity does not survive on either side of the transition. On the string side, however, the response has a distinctive multipolar structure: the Love numbers grow rapidly with $\ell$, and we show analytically that this growth originates from the competition between the multipolar weight of the tidal field and the exponential tail of the winding condensate, with the result that the radial scale probed by the deformability grows linearly with $\ell$ at large multipole number. This provides a direct tidal signature of the extended, surface-less nature of the string star. We discuss the comparison with $\alpha'$-corrected black holes and the limitations of the currently available perturbative results at large $\ell$.
}

\begin{document}

\maketitle

\section{Introduction}
\label{sec:Intro}

Four-dimensional black holes do not deform. Their static tidal Love numbers vanish identically for all multipoles of scalar, electromagnetic, and gravitational perturbations in four-dimensional asymptotically flat spacetime
\cite{Binnington:2009bb,Damour:2009vw,Kol:2011vg,Hui:2020xxx}. In the worldline effective theory these coefficients multiply finite-size operators that no symmetry of the low-energy description forbids, so their vanishing is a naturalness puzzle \cite{Porto:2016zng}: the tidally forced solution must develop no decaying tail whatsoever, a degeneracy with no reason to hold unless something enforces it. The standard resolution invokes a hidden $SL(2,\mathbb{R})$ ``Love symmetry'' of the near-zone perturbation equations \cite{Charalambous:2021mea,Charalambous:2022rre,Hui:2021vcv,Lupsasca:2025pnt,ParraMartinez:2025nvt}, whose highest-weight structure relies on regularity at a sharply defined horizon.\footnote{When the horizon undergoes large quantum fluctuations, as in near-extremal black holes, it becomes deformable \cite{Cano:2026lfl}.}

This structure is delicate, and not simply the imprint of a horizon. Already in $D\geq5$ the symmetry enforces the vanishing of only some of the Schwarzschild-Tangherlini Love numbers, while others are finite or exhibit logarithmic running with the renormalization scale \cite{Kol:2011vg,Hui:2020xxx}, and it is lost altogether when the asymptotics are changed \cite{Emparan:2017qxd,Franzin:2024cah}.\footnote{As far as we are aware, non-zero Love numbers in a four-dimensional black hole were first found in \cite{Emparan:2017qxd}, owing to AdS asymptotics.} Horizonless objects, meanwhile, deform freely, with nothing to protect them \cite{Flanagan:2007ix,Cardoso:2017cfl,Sennett:2017etc,Bianchi:2022qph,Chirenti:2020bas,Giri:2024cks}. A finite tidal response is the generic situation; it is the vanishing that calls for an explanation. Tidal deformability is correspondingly a sharp diagnostic of internal structure, and measurable in principle---one of the cleanest gravitational-wave probes of that structure \cite{Rodriguez:2026iot,Chakraborty:2026qru}.

The question, then, is not whether a horizonless object deforms, but what happens to the fine-tuned rigidity of a black hole as it transforms into an extended object. Neutron and boson stars cannot answer it: they are not black holes in another regime, so setting their Love numbers beside those of a black hole compares unrelated objects. String theory supplies a setting where the question is well posed. As a black hole is made lighter, or the string coupling reduced, its horizon shrinks to the string scale, where the same states are expected to be described at weak coupling by a self-gravitating highly excited string \cite{Susskind:1993ws,Horowitz:1996nw,Damour:1999aw}.\footnote{More precisely, the same states within a band---since a completely smooth transition is only possible for some string theories \cite{Chen:2021dsw}---and up to corrections that scale inversely with the entropy---since the transition can never be perfectly adiabatic \cite{Ceplak:2023afb}.} These are not two objects, but one family of states with two descriptions, matched at the correspondence point.\footnote{Scalarized black holes offer another interesting setting \cite{Silva:2017uqg}, though there the deformation follows from a
modification of the theory, while here we follow one family of states
across different descriptions.} Near the Hagedorn temperature the string-side description is the Horowitz--Polchinski (HP) solution \cite{Horowitz:1997jc}: a condensate of a light Euclidean winding mode, bound by the gravitational potential it sources, with neither horizon nor surface \cite{Chen:2021dsw}. Its tidal response thus continues that of the black hole across the transition, rather than being that of just another horizonless star.

We compute the static tidal Love numbers of the HP string star. At leading order the HP system is a self-consistent Newtonian problem for the winding condensate and the gravitational potential that sets the size of the thermal circle, admitting string star solutions in $D=4,5,6$. We solve the linear response to external tidal fields of multipole order $\ell=2,3,4$ in each dimension, extracting the Love numbers $\lambda_\ell$ from the decaying tail of the induced potential.

The results are non-zero throughout. In $D=4$, where the Schwarzschild black hole has $\lambda_\ell = 0$ for all $\ell$, the string star yields
\begin{equation}
\label{eq:TLN_D4_intro}
    \lambda_2 \simeq 41.5\,,\qquad \lambda_3 \simeq 626\,,\qquad \lambda_4 \simeq 1.68\times10^4\,,
\end{equation}
in units of the characteristic HP radius, $r_{\text{\tiny HP}}$. Nothing in the HP equations enforces a vanishing response, and the same holds for the $\alpha'$-corrected black hole equations, whose Love numbers are likewise finite or running where Einstein gravity predicts none \cite{Charalambous:2024tdj,Katagiri:2024fpn,Cano:2025zyk}. That both sides of the transition deform is therefore expected rather than surprising. What is informative is how much, and how the response is distributed over multipoles. 

We find that the string star response exhibits a particularly distinctive multipolar structure. The Love numbers grow rapidly with $\ell$, and this growth has a simple physical origin: the factor $r^\ell$ in the tidal field pushes the dominant contribution to the induced deformation progressively outward, while the exponential decay of the condensate eventually suppresses it. The resulting saddle point moves to radii $\propto \ell\,r_{\text{\tiny HP}}$ at large $\ell$, so higher multipoles probe progressively farther into the dilute tail of the string star. The multipolar growth is therefore a consequence of its extended, surface-less structure rather than merely a numerical pattern.

It is useful to characterize this growth by a length scale, the `deformability radius',
\begin{equation}
\label{eq:calR_def_intro}
    {\cal R}_\ell \equiv r_{\text{\tiny HP}} \,
    \lambda_\ell^{1/(2\ell+1)},
\end{equation}
which removes the convention dependence associated with the characteristic
radius used to define $\lambda_\ell$, and in particular the growing
sensitivity to this choice at large $\ell$.\footnote{This is in $D=4$; the general definition is given in \eqref{eq:Rell_def}.}
For the numerical results \eqref{eq:TLN_D4_intro} we find
\begin{equation}
    \frac{{\cal R}_2}{r_{\text{\tiny HP}}}\simeq2.11 \, , \qquad
    \frac{{\cal R}_3}{r_{\text{\tiny HP}}}\simeq2.51 \, , \qquad
    \frac{{\cal R}_4}{r_{\text{\tiny HP}}}\simeq2.95 \, .
\end{equation}
Thus ${\cal R}_\ell$ increases steadily over the multipoles studied here. As we show below, this behavior follows from the same saddle-point structure, with ${\cal R}_\ell/r_{\text{\tiny HP}}$ growing linearly with $\ell$ at large multipole number. This mechanism has no direct analogue in the black hole phase, where the response is determined by horizon regularity.

We emphasize that this is a qualitative comparison, not a matching calculation nor a test of the black hole--string correspondence: the HP description is controlled at weak coupling near the Hagedorn temperature, while the $\alpha'$ expansion applies to large black holes, and neither is quantitatively controlled at the correspondence point. Even so, two conclusions stand out. First, the vanishing of the Love numbers in four-dimensional Einstein gravity is fine-tuned: it is lost both under higher-derivative corrections on the black hole side
and in the passage to the string phase. Second, on the string side the multipolar growth exhibits a pattern with a definite physical origin: higher multipoles progressively probe the dilute outer tail of the winding condensate.

The paper is organized as follows. Section~\ref{sec:HP} reviews the HP effective theory and its regime of validity. Section~\ref{sec:UnpertPert} constructs the spherically symmetric background. Section~\ref{sec:TLNsHP} derives the tidal perturbation equations, defines and computes the Love numbers, and then studies the dependence of the deformability radius on the multipole number. Section~\ref{sec:BHstringComparison} compares with black hole results across the transition. Section~\ref{sec:conclusions} summarizes our conclusions and discusses open directions, of which the extension to rotating configurations seems the most immediate. Appendix~\ref{app:numerics} collects the details of the numerical study.

\section{Horowitz--Polchinski string stars}
\label{sec:HP}

Let us review the effective field theory of the static neutral Horowitz--Polchinski (HP) string star: a self-gravitating highly excited string near the Hagedorn temperature that provides the weak-coupling string-side configuration---a string star---expected to connect to a static neutral black hole at the black hole--string transition \cite{Horowitz:1997jc,Chen:2021dsw}. 

The leading HP string star is well-defined without further corrections to the effective theory only in $D=4,5,6$ \cite{Horowitz:1997jc,Chen:2021dsw}. These are the dimensions on which we focus in the numerical computation of the Love numbers. We follow the notation of \cite{Chen:2021dsw,Emparan:2024mbp}, writing 
\begin{align}
    d=D-1 
\end{align}
for the number of spatial dimensions and defining the string mass and length as
\begin{align}
    M_s = 1/\ell_s = 1/\sqrt{\alpha'}\,.
\end{align}

The HP solution describes a string state close to the Hagedorn temperature $1/\beta_H \sim M_s$, and it is naturally studied within the Euclidean thermal path integral. The inverse temperature $\beta$ is implemented by compactifying Euclidean time $\tau$ on a circle, $\tau \sim \tau+\beta$, around which closed strings can wind. The winding sectors are labeled by an integer $n$ and satisfy, schematically, $X^0(\sigma+2\pi) = X^0(\sigma)+n\beta$. Near the Hagedorn temperature the lightest winding modes, $n=\pm 1$, dominate; in the dimensionally reduced $d$-dimensional theory they are described by a complex field $\chi$, the thermal scalar. This field becomes massless at the Hagedorn temperature $\beta=\beta_H$, so for $\beta-\beta_H \ll \ell_s$ it is light compared with the string scale and must be kept explicitly in the low-energy effective theory \cite{Atick:1988si}.

The HP solution is a saddle of the finite-temperature effective theory in which $\chi$ develops a non-zero, spatially localized profile—a ``winding condensate'', a normalizable bound-state wavefunction supported by the self-gravitational potential of the string star itself. We take $\chi$ real and positive, which is possible and convenient for the neutral configurations considered here.

The gravitational degree of freedom playing the leading role is the one controlling the proper length of the Euclidean time circle, which we parametrize as 
\begin{equation}
    \beta_\text{loc}(x) = \beta e^{\varphi(x)}\,,
\end{equation}
with $\beta$ the circle length at asymptotic infinity. Imposing $\varphi \to 0$ at spatial infinity fixes $\beta$ as the asymptotic inverse temperature. For small $\varphi$, $\beta_\text{loc} \approx \beta(1+\varphi)$, so negative $\varphi$ corresponds to a locally smaller Euclidean time circle, equivalently a locally higher redshifted temperature. As the equations of motion below will show, the ``radion field'' $\varphi$ plays the role of the gravitational potential in the effective self-gravitating description of the string star.

The mass of the thermal scalar depends on the local size of the Euclidean time circle. Near the Hagedorn temperature this dependence expands as
\begin{equation}
    m(\varphi)^2 = m_\infty^2 + {\mathcal{M}_s}^2 \varphi + O(\varphi^2) \, ,
\end{equation}
where the asymptotic scalar mass is
\begin{equation}
    m_\infty^2 = {\mathcal{M}_s}^2 \Delta_\beta 
    \,,
\end{equation}
and we have conveniently introduced
\begin{equation}
    {\mathcal{M}_s^2} = \frac{\kappa}{\alpha'} = \kappa M_s^2 \, , \qquad \Delta_\beta = \frac{\beta-\beta_H}{\beta_H} \, .
\end{equation}
The constant $\kappa$ is an $O(1)$ number set by the string theory: $\kappa=8$ for the bosonic string, $\kappa=4$ for type II, and $\kappa=4\sqrt{2}$ for heterotic \cite{Chen:2021dsw}. Its value will not matter in what follows, and it will always be absorbed into ${\mathcal{M}_s}$.

We consider configurations below the Hagedorn temperature throughout, so $\Delta_\beta>0$. The condition
\begin{equation}
\label{eq:db_small}
    \Delta_\beta \ll 1 \, , 
\end{equation}
states that the system is near the Hagedorn temperature and that the thermal scalar is light compared with the string scale,
\begin{equation}
    m_\infty^2 \ll{ \mathcal{M}_s}^2 \sim M_s^2 \, .
\end{equation}

After reduction on the Euclidean time circle, the low-energy effective theory in the $d$ non-compact directions contains the spatial metric $g_{ab}$, the dilaton $\Phi$, the radion $\varphi$, and the thermal scalar $\chi$, with action \cite{Horowitz:1997jc,Chen:2021dsw}
\begin{equation}
    I = \frac{1}{16 \pi G_N} \, \int d^d x \, \sqrt{g} \, e^{-2 \Phi} \, \left( -\mathcal{R} - 4 \, (\nabla \Phi)^2 + (\nabla \varphi)^2 + |\nabla \chi|^2 + m(\varphi)^2 \, |\chi|^2 + \cdots \right) \, .
\end{equation}
Here $\mathcal{R}$ is the Ricci scalar of $g_{ab}$ and $G_N$ is Newton's constant in $D$ spacetime dimensions. The dots denote terms subleading in the HP regime—higher powers of the fields, higher-derivative and string-loop corrections—which we consistently neglect.

The coupling between $\chi$ and $\varphi$ is the essential ingredient of the HP construction. Through the expansion of the thermal scalar mass, the action contains the interaction ${\mathcal{M}_s}^2 \varphi |\chi|^2$, so the local size of the Euclidean time circle sets the effective mass of the winding mode. Its interpretation will become transparent once we derive the equations of motion.

\subsection{Equations of motion}
 \label{subsec:EOMs}

At leading order in the near-Hagedorn expansion, the dominant backreaction is the coupling between $\chi$ and $\varphi$, while the dilaton $\Phi$ and spatial metric $g_{ab}$ can be held fixed. Setting $g_{ab} = \delta_{ab}$ and $\Phi = \text{constant}$ and varying with respect to $\chi$ and $\varphi$ gives
\begin{equation}
\label{eq:EOM1}
    \begin{cases}
      \nabla^2 \chi = {\mathcal{M}_s^2}\left(\Delta_\beta + \varphi \right) \chi\,,\\
      \nabla^2 \varphi = \frac{1}{2} \, {\mathcal{M}_s^2}\,\chi^2\,,
    \end{cases}
\end{equation}
where $\nabla^2$ is the flat Laplacian in $d$ spatial dimensions. The interpretation is simple. The first equation makes $\chi$ a bound-state wavefunction in the effective potential set by the local size of the thermal circle: since $\varphi$ can be negative in the interior, the combination $\Delta_\beta+\varphi$ is reduced relative to its asymptotic value, allowing a localized profile of $\chi$. The second equation shows that this same profile sources $\varphi$. The HP string star is thus a self-consistent bound state—the winding condensate sources the potential that traps it.

The total mass $M$ of the condensate follows either from the thermodynamic analysis of the Euclidean action or, equivalently, from the asymptotic fall-off of $\varphi$. Combining eqs. (2.14) and (2.18) of \cite{Chen:2021dsw},
\begin{equation}
\label{eq:HPmass}
    M = \frac{\mathcal{M}_s^2}{16 \pi G_N} \int d^d x \, \chi(x)^2 \, ,
\end{equation}
so $\chi^2$ is proportional to the mass density $\rho$,
\begin{equation}
\label{eq:rho_chi_original}
    \rho = \frac{\mathcal{M}_s^2}{16 \pi G_N} \, \chi^2 \, .
\end{equation}
This is the precise sense in which the winding condensate describes the spatial distribution of the highly excited string.

To make the relation between $\varphi$ and the Newtonian potential $U$ explicit, we define the latter so that the $d$-dimensional Poisson equation reads\footnote{For $D>4$, our normalization \eqref{eq:Poisson_convention} of the Newtonian potential differs by a dimension-dependent constant from the standard normalization in $D$-dimensional Einstein gravity. This also fixes the normalization of the Love numbers used below; in $D=4$ the two normalizations coincide.}
\begin{equation}
\label{eq:Poisson_convention}
    \nabla^2 U = \omega_{d-1} \, G_N \, \rho \, ,
\end{equation}
with $\omega_{d-1}$ the area of the unit $(d-1)$-sphere,
\begin{equation}
\label{eq:omega_unit_sphere}
    \omega_{d-1} = \frac{2 \pi^{d/2}}{\Gamma(d/2)} \, .
\end{equation}
Combining \eqref{eq:EOM1}, \eqref{eq:rho_chi_original} and \eqref{eq:Poisson_convention},
\begin{equation}
\label{eq:U_varphi_original}
    U = \frac{\omega_{d-1}}{8\pi} \, \varphi \, ,
\end{equation}
so, although $\varphi$ was introduced geometrically as the fluctuation of the Euclidean thermal circle, the equations of motion show that it plays the role of the Newtonian gravitational potential. In particular, for a spherically symmetric configuration of total mass $M$,
\begin{equation}
\label{eq:varphi_mass_asymptotic}
    \varphi(r) = - \frac{8\pi}{(d-2)\omega_{d-1}} \, \frac{G_N M}{r^{d-2}} + O(r^{1-d}) \, .
\end{equation}

To obtain the solutions and study the tidal deformability it is convenient to remove all dimensionful parameters. From now on we measure lengths in string units and shift the potential by its asymptotic value, defining
\begin{equation}
\label{eq:tilde_variables_def}
    \tilde{x} = \mathcal{M}_s \, x \, , \qquad \tilde{\chi}(\tilde{x}) = \frac{\chi(x)}{\sqrt{2}} \, , \qquad \tilde{\varphi}(\tilde{x}) = \varphi(x) + \Delta_\beta \, .
\end{equation}
In these variables \eqref{eq:EOM1} become
\begin{equation}
\label{eq:UEOM}
    \begin{cases}
      \tilde{\nabla}^2 \tilde{\chi} = \tilde{\varphi} \, \tilde{\chi}\,,\\
      \tilde{\nabla}^2 \tilde{\varphi} = \tilde{\chi}^2\,.
    \end{cases}
\end{equation}
This is the universal, parameter-free form of the HP equations: all explicit dependence on ${\mathcal{M}_s}$ and $\Delta_\beta$ has disappeared. The physical value of $\Delta_\beta$ is recovered from the asymptotics of the shifted potential, whose boundary condition is now not $\tilde{\varphi} \to 0$ but
\begin{equation}
\label{eq:varphi_infinity_delta_beta}
    \tilde{\varphi}(\tilde{r}) \to \Delta_\beta \quad \text{as} \quad \tilde{r} \to \infty \, .
\end{equation}
Thus $\Delta_\beta$ is determined dynamically by the asymptotic value of the solution in the universal variables.

In the same variables the mass density and Newtonian potential are
\begin{equation}
\label{eq:tilde_rho_U}
    \tilde{\rho} = \frac{{\mathcal{M}_s^{2-d}}}{8 \pi G_N} \, \tilde{\chi}^2 \, , \qquad \tilde{U} = \frac{\omega_{d-1} \, \mathcal{M}_s^{2-d}}{8 \pi} \left(\tilde{\varphi}-\Delta_\beta\right) \, ,
\end{equation}
and satisfy
\begin{equation}
    M = \int d^d \tilde{x} \, \tilde{\rho} \, , \qquad \tilde{\nabla}^2 \tilde{U} = \omega_{d-1} \, G_N \, \tilde{\rho} \, .
\end{equation}
Note that $\tilde{\rho}(\tilde{x})$ and $\tilde{U}(\tilde{x})$ are related to their counterparts in the original variables by a simple dimensionful rescaling,
\begin{equation}
\label{eq:rho_U_tilde_original}
    \tilde{\rho}(\tilde{x}) = \mathcal{M}_s^{-d} \, \rho\big(x(\tilde{x}) \big) \, , \qquad \tilde{U}(\tilde{x}) = \mathcal{M}_s^{2-d} \, U\big(x(\tilde{x}) \big) \, .
\end{equation}

\subsection{Regime of validity and scaling symmetry}
\label{subsec:validity_scaling}

The leading HP effective theory is valid only in a finite parameter window. First, the thermal scalar must be light compared with the string scale, requiring \eqref{eq:db_small}, and the field amplitude must remain small, so that higher powers of the fields in the action can be neglected. If either $\Delta_\beta$ or the condensate amplitude becomes of order one, the leading theory breaks down and higher stringy corrections take over.

Second, the system cannot be taken arbitrarily close to the Hagedorn point at fixed string coupling $g_s$: there the thermal scalar becomes very light and quantum fluctuations of the condensate grow large. Validity of the classical HP saddle requires \cite{Chen:2021dsw}
\begin{equation}
\label{eq:Debebounds}
    g_s^{\frac{4}{7-D}} \lesssim \Delta_\beta \ll 1 \, .
\end{equation}
Below the lower end, the saddle has action of order one and the right description is instead a weakly interacting long string, with Hagedorn entropy $S \approx \beta_H M$ and random-walk size of order $\sqrt{S}$ in string units. At the upper end, where $\Delta_\beta$ is of order one, the effective theory again fails; this is where the solution is expected to connect to the black hole. As $D=7$ is approached from below the window shrinks, and for $D\geq 7$ corrections to the leading theory are needed \cite{Balthazar:2022hno}.

The universal equations \eqref{eq:UEOM} have a simple scaling symmetry: if $\tilde{\chi}(\tilde{x})$ and $\tilde{\varphi}(\tilde{x})$ solve them, so does
\begin{equation}
\label{eq:scaling}
    \left(\tilde{x}^i,\tilde{\chi},\tilde{\varphi}\right) \mapsto \left(\lambda^{-1/2}\tilde{x}^i,\lambda \tilde{\chi},\lambda \tilde{\varphi}\right) \, .
\end{equation}
Under this map $\tilde{\nabla}^2 \mapsto \lambda \tilde{\nabla}^2$, and both sides of each equation scale identically. Since $\tilde{\varphi}(\infty)=\Delta_\beta$, it also sends $\Delta_\beta \mapsto \lambda \Delta_\beta$. A single solution therefore generates a one-parameter family by rescaling its amplitude, size, and temperature---a property that string stars share with classical black holes. The symmetry is exact only in the leading HP system; it is broken by the higher-order terms in the action and by the quantum effects that set the window \eqref{eq:Debebounds}. 

One consequence of \eqref{eq:scaling} is that $\tilde{x}^i\sqrt{\Delta_\beta}$ is scale-invariant, so in the dimensionless coordinates of \eqref{eq:UEOM} the characteristic size of the string star is $\sim 1/\sqrt{\Delta_\beta}$. As $\beta \to \beta_H$, i.e. $\Delta_\beta \to 0$, the star becomes increasingly large and diffuse. The size of the string star can be more precisely defined from the asymptotic fall-off of the condensate, and we will see below that this estimate is thus reproduced (cf.~\eqref{eq:rb}).

The same symmetry lets us fix the overall condensate amplitude in the numerical construction. We choose
\begin{equation}
    \tilde{\chi}(0)=1
\end{equation}
for convenience. The HP effective theory requires the physical amplitude to be small, but this poses no conflict: the scaling symmetry can rescale any solution to smaller physical amplitude, so fixing $\tilde{\chi}(0)=1$ simply selects a representative of the scaling orbit. The dimensionless Love numbers that we define below are invariant under these rescalings.

\section{Unperturbed solution}
\label{sec:UnpertPert}

For the unperturbed static neutral string star the fields depend only on $\tilde{r}$, and \eqref{eq:UEOM} becomes
\begin{equation}
\label{eq:EOMUnpert}
    \begin{cases}
      \tilde{\chi}'' + \frac{d-1}{\tilde{r}} \, \tilde{\chi}' = \tilde{\varphi} \, \tilde{\chi}\,,\\
      \tilde{\varphi}'' + \frac{d-1}{\tilde{r}} \, \tilde{\varphi}' = \tilde{\chi}^2\,,
    \end{cases}
\end{equation}
with a prime denoting $\tilde{r}$-differentiation.

Regularity at the origin requires $\{\tilde{\chi}'(0)=0, \tilde{\varphi}'(0)=0\}$, and normalizability requires $\tilde{\chi}$ to decay at infinity. As shown in Section~\ref{sec:HP}, $\tilde{\varphi}$ approaches $\Delta_\beta$ there, so
\begin{equation}
\label{eq:chvplim}
    \tilde{\chi} \rightarrow 0 \, , \qquad \tilde{\varphi} \rightarrow \Delta_\beta \, , \qquad \text{for } \tilde{r} \rightarrow \infty \, .
\end{equation}
Fixing the condensate at the origin via the scaling symmetry \eqref{eq:scaling}, the boundary conditions for the numerical construction are
\begin{equation}
\label{eq:BCsUnpert}
    \tilde{\chi}(0)=1 \, , \qquad \tilde{\chi}'(0)=0 \, , \qquad \tilde{\varphi}'(0)=0 \, , \qquad \tilde{\chi}(\infty)=0 \, .
\end{equation}
The value $\tilde{\varphi}(0)$ cannot be specified independently; it is fixed by requiring a regular, normalizable state. We focus on the ground state, for which $\tilde{\chi}$ has no nodes.

The behavior near the origin follows from \eqref{eq:EOMUnpert}: regularity means a smooth spherically symmetric function expands in even powers of $\tilde{r}$, and using $\tilde{\nabla}^2(\tilde{r}^2) = 2d$ we have
\begin{equation}
\label{eq:chi_phi_small_r}
    \tilde{\chi}(\tilde{r}) =
    1 + \frac{\tilde{\varphi}_0}{2d} \, \tilde{r}^2 + O(\tilde{r}^4) \, , \qquad \tilde{\varphi}(\tilde{r}) =
    \tilde{\varphi}_0 + \frac{1}{2d} \, \tilde{r}^2 + O(\tilde{r}^4) \, , \qquad \text{for } \tilde{r} \to 0 \, ,
\end{equation}
with $\tilde{\varphi}_0 = \tilde{\varphi}(0)$.
At large radius the condensate is exponentially small and the second equation reduces to the radial Laplace equation, giving
\begin{equation}
\label{eq:vpNI}
    \tilde{\varphi}(\tilde{r}) \sim \Delta_\beta + \frac{c_\varphi}{\tilde{r}^{d-2}} \qquad \text{for } \tilde{r} \to \infty \, ,
\end{equation}
so that
\begin{equation}
\label{eq:cinfvp}
    c_\varphi
    =
    - \frac{1}{d-2}
    \lim_{\tilde{r}\rightarrow\infty}
    \tilde{r}^{d-1}\tilde{\varphi}'(\tilde{r}) \, .
\end{equation}
Then, in the numerical integration we extract $\Delta_\beta = \tilde{\varphi}(\infty)$ at some large (finite) $\tilde{r}$ from
\begin{equation}
\label{eq:vpinf}
    \Delta_\beta
    =
    \lim_{\tilde{r}\rightarrow\infty}
    \left(
        \tilde{\varphi}(\tilde{r}) - \frac{c_\varphi}{\tilde{r}^{d-2}}
    \right) \, .
\end{equation}
Since the source $\tilde{\chi}^2$ is positive, $c_\varphi<0$ and $\tilde{\varphi}$ approaches $\Delta_\beta$ from below.
On the other hand, the first equation in \eqref{eq:EOMUnpert} reduces to
\begin{equation}
    \tilde{\chi}''+\frac{d-1}{\tilde{r}} \, \tilde{\chi}' \approx \left(\Delta_\beta + \frac{c_\varphi}{\tilde{r}^{d-2}} \right) \, \tilde{\chi} \, ,
\end{equation}
whose normalizable solution behaves as
\begin{equation}
\label{eq:chNI}
    \tilde{\chi}(\tilde{r}) \sim c_\chi \, \frac{e^{-\sqrt{\Delta_\beta} \, \tilde{r}}}{\tilde{r}^{\frac{d-1}{2}+\nu}}\,,\qquad \nu= \frac{c_\varphi}{2\sqrt{\Delta_\beta}}\delta_{d,3} \qquad \text{for } \tilde{r} \to \infty \,,
\end{equation}
where we have taken into account that the Coulomb tail of the potential is asymptotically negligible when $d>3$ but not in $d=3$.  The numerical solution for $d=3$ yields
\begin{equation}\label{eq:cphinu}
    c_\varphi \simeq -3.47\,,\qquad \nu \simeq -1.75\,.
\end{equation}
Notice that $1+\nu <0$ implies that in $d=3$ the power-law factor slows the decay relative to a pure exponential, whereas in $d>3$ it makes the decay faster.

We solve \eqref{eq:EOMUnpert} with \eqref{eq:BCsUnpert} by shooting on $\tilde{\varphi}_0 = \tilde{\varphi}(0)$, selecting the unique nodeless ground state that decays at infinity. The values of $\tilde{\varphi}_0$, $\Delta_\beta$, and $\tilde{r}_{\text{\tiny HP}}$ (cf.~\eqref{eq:rb}) in $D=4,5,6$ are given in Table~\ref{tab:HPBackgroundData}.

\begin{table}[t!]
\centering
\small
\begin{tabular}{|c||c|c|c|}
    \hline
    $D$ & $\tilde{\varphi}_0$ & $\Delta_\beta=\tilde{\varphi}(\infty)$ & $\tilde{r}_{\text{\tiny HP}}=\Delta_\beta^{-1/2}$ \\
    \hline\hline
    $4$ & $-0.91858$ & $0.97896$ & $1.01069$ \\
    \hline
    $5$ & $-0.96827$ & $0.28232$ & $1.88206$ \\
    \hline
    $6$ & $-0.99196$ & $0.07133$ & $3.74432$ \\
    \hline
\end{tabular}
\caption{\small Numerical data for the unperturbed HP string star in $D=4,5,6$. Here $\tilde{\varphi}_0$ is the parameter selected by the shooting procedure, $\Delta_\beta$ is extracted from the large-radius asymptotics of $\tilde{\varphi}$, and $\tilde{r}_{\text{\tiny HP}}=\Delta_\beta^{-1/2}$ is the characteristic radius, cf.~\eqref{eq:rb}. The extraction of $\Delta_\beta$ is stable under variations of the large-radius sampling window at a relative level of approximately
$3.5\times10^{-11}$ or better; see Appendix~\ref{app:numerics}.}
\label{tab:HPBackgroundData}
\end{table}

\begin{figure}[t!]
    \centering
    \includegraphics[width=0.475\textwidth]{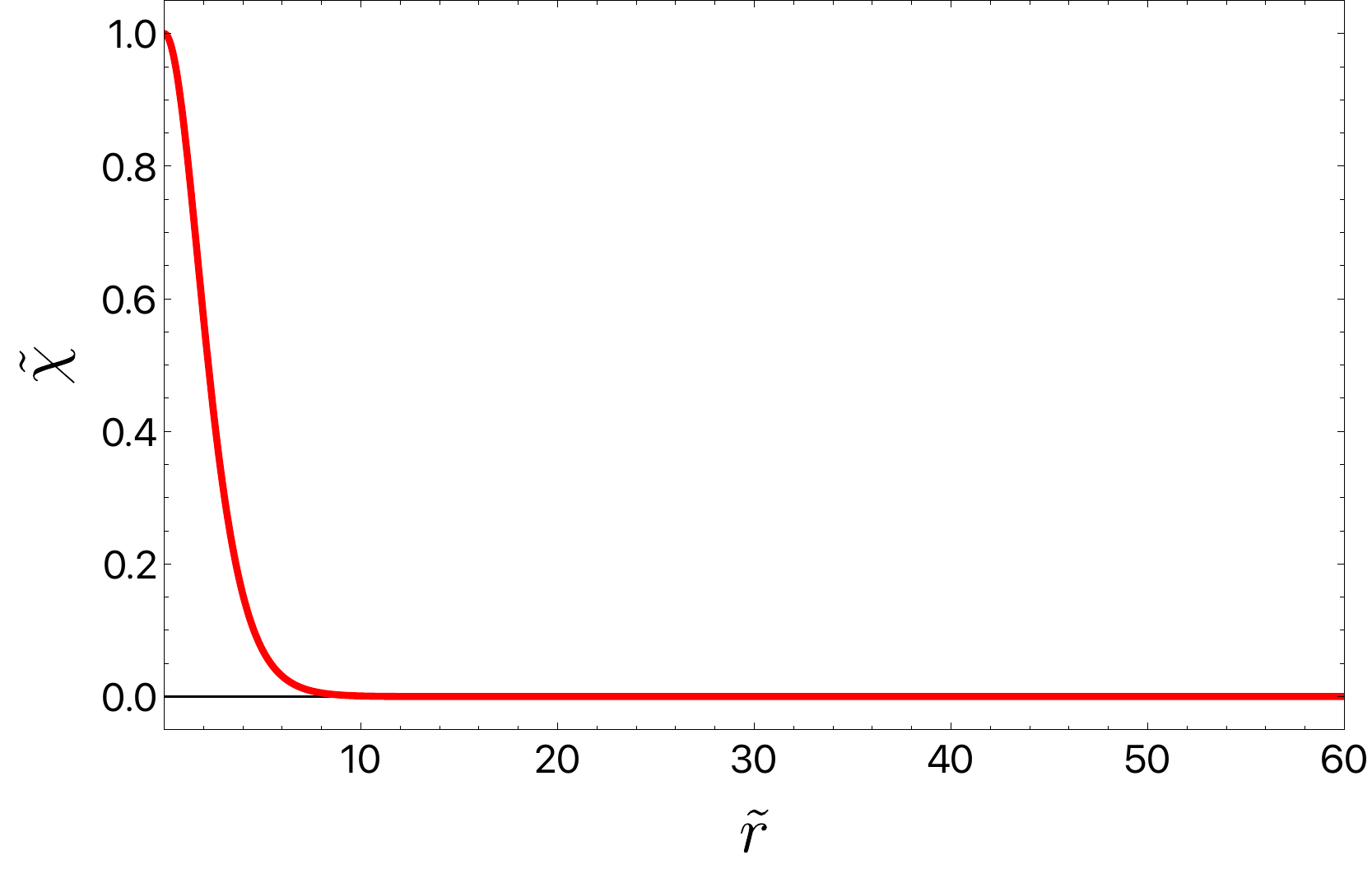}\qquad\includegraphics[width=0.475\textwidth]{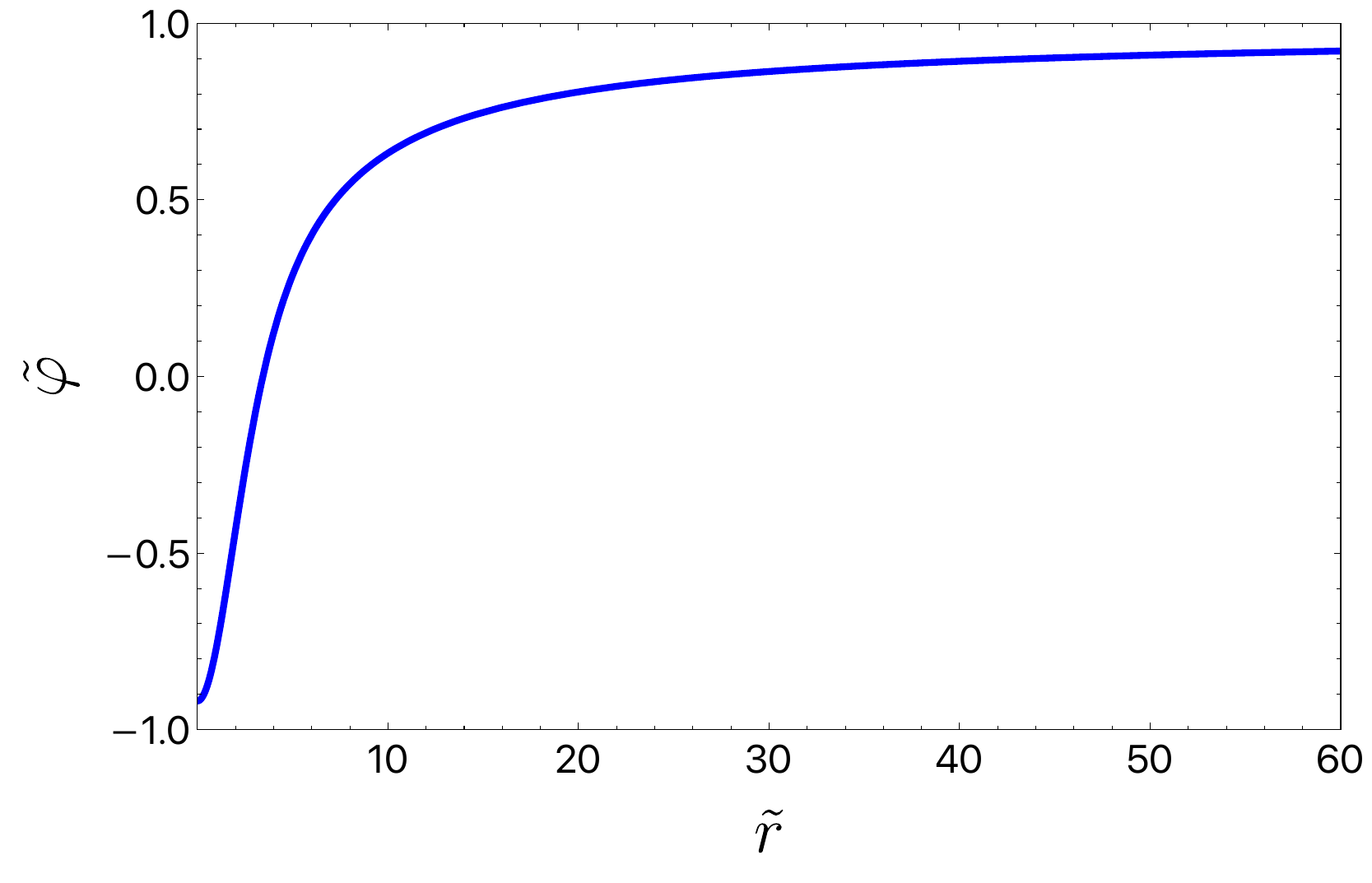}
    \caption{\small Unperturbed HP string star in $D=4$. The rescaled winding condensate $\tilde{\chi}$ (left) and shifted potential $\tilde{\varphi}$ (right) are shown as functions of the dimensionless radius $\tilde{r}$, and are obtained by solving \eqref{eq:EOMUnpert} with boundary conditions \eqref{eq:BCsUnpert} for the normalizable ground state. The rescaled variables $\tilde{r}, \tilde{\chi}, \tilde{\varphi}$ relate to the original $r,\chi,\varphi$ via \eqref{eq:tilde_variables_def}, and to the mass density $\rho$ and Newtonian potential $U$ via \eqref{eq:rho_chi_original}, \eqref{eq:U_varphi_original}, \eqref{eq:tilde_rho_U}.}
    \label{fig:chi_phi}
\end{figure}

Figure~\ref{fig:chi_phi} shows the numerical profiles in $D=4$. Those in $D=5,6$ are qualitatively similar: $\tilde{\chi}$ decreases monotonically from $1$ to zero, while $\tilde{\varphi}$ starts negative and approaches a positive constant.

The exponential decay \eqref{eq:chNI} motivates defining the string star radius
\begin{equation}
\label{eq:rb}
    \tilde{r}_{\text{\tiny HP}} = \frac{1}{\sqrt{\Delta_\beta}} = \sqrt{\frac{\beta_H}{\beta-\beta_H}} \, .
\end{equation}
The physical radius, which we denote without a tilde, is
$r_{\text{\tiny HP}}=\tilde r_{\text{\tiny HP}}/\mathcal{M}_s$. 

The tabulated values show that $\tilde{r}_{\text{\tiny HP}}$ grows with spacetime dimension: the HP solutions become less localized as $D$ approaches the critical value $D=7$. For $D \geq 7$ the gravitational interaction is too short-ranged to bind the condensate and the leading equations \eqref{eq:EOMUnpert} have no normalizable string star solutions with $\Delta_\beta > 0$ \cite{Horowitz:1997jc,Chen:2021dsw}. Thus we restrict ourselves to $D=4,5,6$.\footnote{See \cite{Balthazar:2022hno,Bedroya:2024igb} for the existence of solutions in $D \geq 7$ including higher-order terms in the effective action.} Notice from Figure~\ref{fig:chi_phi} that, with this definition, most of the condensate lies outside $\tilde{r}_{\text{\tiny HP}}$: it is the e-folding length of the tail, not a radius enclosing the bulk of the star.

Finally, although the string star is a classical condensate, it has an entropy as a result of the dependence of its radial profile on $\beta$. This entropy reproduces the degeneracy of string states at a given mass \cite{Chen:2021dsw}.

\section{Love numbers of the HP string star}
\label{sec:TLNsHP}

\subsection{Tidal perturbation equations}\label{subsec:tidalpert}

An external static tidal potential deforms both the condensate and the self-generated potential. In the dimensionless variables, $\tilde{U}$ is related to $\tilde{\varphi}$ by \eqref{eq:tilde_rho_U}, so we write the external tidal potential as
\begin{equation}
\label{eq:Vext}
    V_\text{ext}(\tilde{x}) = \frac{\omega_{d-1} \, \mathcal{M}_s^{2-d}}{8 \pi} \, \eta(\tilde{x}) \, ,
\end{equation}
with $\eta(\tilde{x})$ a dimensionless driving field. Since its source lies far from the string star, $\eta$ obeys Laplace's equation in the region occupied by the star,
\begin{equation}
\label{eq:etLaplace}
    \tilde{\nabla}^2 \eta = 0 \, .
\end{equation}
We denote the resulting first-order perturbations of the condensate and the self-generated potential by $\delta\tilde{\chi}$ and $\delta\tilde{\varphi}$, respectively, so that the perturbed fields are $\tilde{\chi}+\delta\tilde{\chi}$ and $\tilde{\varphi}+\delta\tilde{\varphi}$.
The field entering the condensate equation is the total potential
\begin{equation}
    \tilde{\varphi}_\text{tot}
    =
    \tilde{\varphi} + (\delta\tilde{\varphi} + \eta) \, .
\end{equation}
The distinction between the two perturbations matters: $\eta$ is the applied tidal field, while $\delta\tilde{\varphi}$ is the induced response of the string star.

To first order around the spherically symmetric background, \eqref{eq:UEOM} becomes
\begin{equation}
\label{eq:EOMPert}
    \begin{cases}
        \tilde{\nabla}^2 \delta\tilde{\chi}
        =
        \tilde{\varphi} \, \delta\tilde{\chi}
        + \tilde{\chi} \, \left(\delta\tilde{\varphi} + \eta\right)\\
        \tilde{\nabla}^2 \delta\tilde{\varphi}
        =
        2 \, \tilde{\chi} \, \delta\tilde{\chi}
    \end{cases}
\end{equation}
where \eqref{eq:etLaplace} was used in the second equation. The first says the external field shifts the effective mass of the condensate and so deforms its density profile; the second says this induced density change sources the response potential $\delta\tilde{\varphi}$.

The tidal Love numbers of the HP string star are extracted from the solutions of \eqref{eq:EOMPert} entirely within the effective Newtonian description above: the response field is the induced perturbation $\delta\tilde{U}$ of the Newtonian potential of the string star (cf.~\eqref{eq:tilde_rho_U}--\eqref{eq:rho_U_tilde_original}), and the source is the applied tidal potential $V_\text{ext}$ (cf.~\eqref{eq:Vext}).\footnote{\label{radion_caveat}Both are pure radion perturbations. On the source side, an external mass would also generate tidal components of the spatial metric and of the reduced dilaton $\Phi$, but these reach the condensate only through the Laplacian and are suppressed by one further power of $\Delta_\beta$ relative to $\eta$. On the response side, the radion perturbation uplifts to a fixed combination of the $D$-dimensional graviton and dilaton rather than to a purely
gravitational mode; separating the two channels lies beyond the leading HP truncation used here.}

\subsection{Definition} Given the spherical symmetry of the background, we decompose the tidal field and the response in scalar harmonics $Y_{\ell m}(\Omega)$ on $S^{d-1}$, where $\ell=0,1,2,\ldots$ is the degree, $m$ collects the degeneracy labels, and $\Omega=(\theta_1,\ldots,\theta_{d-1})$ are the angular coordinates. They satisfy
\begin{equation}
    \nabla^2_{S^{d-1}} Y_{\ell m}(\Omega)
    =
    - \ell(\ell+d-2) \, Y_{\ell m}(\Omega) \, ,
\end{equation}
with $\nabla^2_{S^{d-1}}$ the Laplacian on $S^{d-1}$.

Since the external tidal potential is harmonic, its regular growing part is
\begin{equation}
\label{eq:VextSH}
    V_\text{ext}(\tilde{r},\Omega)
    =
    \sum_{\ell=2}^{\infty}\sum_{m}
    A_{\ell m} \,
    \tilde{r}^{\ell} \,
    Y_{\ell m}(\Omega) \, .
\end{equation}
The sum starts at $\ell=2$: monopole perturbations change the total mass or temperature of the configuration, and dipole perturbations translate the center of mass without deforming it.

The external field perturbs the mass density, and hence the Newtonian potential. Writing
\begin{equation}
    \delta\tilde{\rho}(\tilde{r},\Omega)
    =
    \sum_{\ell=2}^{\infty}\sum_{m}
    \delta\tilde{\rho}_{\ell m}(\tilde{r}) \,
    Y_{\ell m}(\Omega) \, ,
\end{equation}
the induced multipole moments are
\begin{equation}
\label{eq:Ilm}
    B_{\ell m}
    =
    \int_0^\infty d\tilde{r} \,
    \tilde{r}^{\ell+d-1} \,
    \delta\tilde{\rho}_{\ell m}(\tilde{r}) \, ,
\end{equation}
and the Poisson convention \eqref{eq:Poisson_convention} gives the asymptotic induced potential
\begin{equation}
\label{eq:deltaUinfty}
    \delta\tilde{U}(\tilde{r},\Omega)
    \sim
    -
    \omega_{d-1} \, G_N
    \sum_{\ell=2}^{\infty}\sum_{m}
    \frac{B_{\ell m}}{2\ell+d-2} \,
    \frac{Y_{\ell m}(\Omega)}{\tilde{r}^{\ell+d-2}} \qquad \text{for } \tilde{r} \to \infty \, .
\end{equation}

The Newtonian tidal Love numbers are the linear-response coefficients relating the induced moments to the source coefficients:
\begin{equation}
\label{eq:TLNsdef}
    \lambda_\ell = -\frac{G_N}{2 \, \tilde{r}_{\text{\tiny HP}}^{2\ell+d-2}} \, \frac{B_{\ell m}}{A_{\ell m}} \, .
\end{equation}
The factor $\tilde{r}_{\text{\tiny HP}}^{2\ell+d-2}/G_N$ renders $\lambda_\ell$ dimensionless, and the factor of $2$ matches the standard four-dimensional Newtonian convention \cite{Binnington:2009bb,Poisson_2014}; $\tilde{r}_{\text{\tiny HP}}$ is the HP radius \eqref{eq:rb}. Because the string star does not have a sharp surface, $\tilde{r}_{\text{\tiny HP}}$ is a characteristic radius, not a boundary. Thus the Love numbers we compute are response coefficients read off from the asymptotic gravitational field, and not ``surficial Love numbers''.

Spherical symmetry keeps different harmonics from mixing, and it implies that the response-to-source ratio depends only on $\ell$, not on $m$. We therefore suppress the $m$ label and work with one fixed multipole $\ell$ at a time.

\subsection{Radial perturbation equations} In terms of the driving field $\eta$, the relation \eqref{eq:Vext} gives
\begin{equation}
    A_\ell
    =
    \frac{\omega_{d-1} \, \mathcal{M}_s^{2-d}}{8\pi} \, f_\ell \, ,
\end{equation}
where the $\ell$-th component of the driving field is
\begin{equation}
\label{eq:eta_ell_general}
    \eta(\tilde{r},\Omega)
    =
    f_\ell \, \tilde{r}^{\ell} \, Y_\ell(\Omega) \, .
\end{equation}
The perturbation equations are linear, so the solution is proportional to $f_\ell$. We conveniently set
\begin{equation}
\label{eq:fell_choice}
    f_\ell = 1 \, , \qquad A_\ell = \frac{\omega_{d-1} \, \mathcal{M}_s^{2-d}}{8\pi}
\end{equation}
in the numerics, independently of $\ell$—i.e. we take the external tidal potential \eqref{eq:Vext} to be the multipolar perturbation
\begin{equation}
\label{eq:Vext_choice}
    V_\text{ext}(\tilde{r},\Omega) = \frac{\omega_{d-1} \, \mathcal{M}_s^{2-d}}{8 \pi} \, \tilde{r}^{\ell} \, Y_\ell(\Omega) \, .
\end{equation}

Decomposing the perturbations as
\begin{equation}
    \delta\tilde{\chi}(\tilde{r},\Omega)
    =
    \delta\tilde{\chi}_\ell(\tilde{r}) \, Y_\ell(\Omega) \, ,
    \qquad
    \delta\tilde{\varphi}(\tilde{r},\Omega)
    =
    \delta\tilde{\varphi}_\ell(\tilde{r}) \, Y_\ell(\Omega) \, ,
\end{equation}
and using
\begin{equation}
    \tilde{\nabla}^2\big(
        F(\tilde{r}) \, Y_\ell(\Omega)
    \big)
    =
    \left[
        F''(\tilde{r}) +\frac{d-1}{\tilde{r}} \, F'(\tilde{r})
        -
        \frac{\ell(\ell+d-2)}{\tilde{r}^2} \, F(\tilde{r})
    \right] \,
    Y_\ell(\Omega) \, ,
\end{equation}
the linearized equations \eqref{eq:EOMPert} become 
\begin{equation}
\label{eq:EOMPertSH}
    \begin{cases}
        \tilde{r}^2 \, \delta\tilde{\chi}_\ell''
        +
        (d-1) \, \tilde{r} \, \delta\tilde{\chi}_\ell'
        -
        \left[\tilde{r}^2 \, \tilde{\varphi} + \ell(\ell+d-2) \right] \, \delta\tilde{\chi}_\ell
        = \tilde{r}^2 \, \tilde{\chi} \, \left(\delta\tilde{\varphi}_\ell+\tilde{r}^{\ell}\right)\\
        \tilde{r}^2 \, \delta\tilde{\varphi}_\ell''
        +
        (d-1) \, \tilde{r} \, \delta\tilde{\varphi}_\ell'
        -
        \ell(\ell+d-2) \, \delta\tilde{\varphi}_\ell
        =
        2 \, \tilde{r}^2 \, \tilde{\chi} \, \delta\tilde{\chi}_\ell
    \end{cases}
\end{equation}
for each $\ell=2,3,\ldots$.

\subsection{Boundary conditions and asymptotic response} The inner boundary condition is regularity at the origin, as in the unperturbed case. For the $\ell$-th multipole, smoothness at $\tilde{r}=0$ requires
\begin{equation}
\label{eq:PertOrigin}
    \delta\tilde{\chi}_\ell(\tilde{r})
    =
    a_\ell \, \tilde{r}^{\ell}
    +
    O(\tilde{r}^{\ell+2}) \, ,
    \qquad
    \delta\tilde{\varphi}_\ell(\tilde{r})
    =
    b_\ell \, \tilde{r}^{\ell}
    +
    O(\tilde{r}^{\ell+2}) \, ,
    \qquad
    \text{for } \tilde{r}\to0 \, ,
\end{equation}
which for $\ell\geq2$ gives $\delta\tilde{\chi}_\ell(0)=\delta\tilde{\varphi}_\ell(0)=0$ and $\delta\tilde{\chi}'_\ell(0)=\delta\tilde{\varphi}'_\ell(0)=0$.

At large $\tilde r$, the first equation in \eqref{eq:EOMPertSH}
becomes
\begin{equation}
\label{eq:approx_eq_delta_chi}
    \delta\tilde{\chi}_\ell''
    +
    \frac{d-1}{\tilde r}\delta\tilde{\chi}_\ell'
    -
    \left[
        \Delta_\beta
        +
        \frac{c_\varphi}{\tilde r^{d-2}}
        +
        \frac{\ell(\ell+d-2)}{\tilde r^2}
    \right]
    \delta\tilde{\chi}_\ell
    \simeq
    \tilde r^\ell\tilde\chi \, .
\end{equation}
Here we have neglected
$\tilde\chi\,\delta\tilde{\varphi}_\ell$ relative to $\tilde r^\ell\tilde\chi$. At fixed $\ell$ the centrifugal term is also negligible as $\tilde{r}\to\infty$, but it will play a role later when we investigate the large-$\ell$ response, which is controlled by radii of order $\tilde r\sim\ell/\sqrt{\Delta_\beta}$ (cf.~\eqref{eq:saddle}), where $\ell^2/\tilde r^2$ must be retained. Lastly, while negligible in $d>3$, the Coulomb term is needed in $d=3$: although it is pointwise subleading relative to $\Delta_\beta$, it changes the leading power-law prefactor of the exponentially decaying condensate, as noted in \eqref{eq:chNI}.

Using \eqref{eq:chNI},
\begin{equation}
    \tilde r^\ell\tilde\chi
    \simeq
    c_\chi\tilde r^\ell
    \frac{e^{-\sqrt{\Delta_\beta}\tilde r}}
    {\tilde r^{(d-1)/2+\nu}} \, .
\end{equation}
The source has the same exponential dependence as the decaying
homogeneous solution, so the forcing is resonant and the particular
solution acquires one additional power of $\tilde r$,
\begin{equation}
\label{eq:dechi_asymptotic}
    \delta\tilde{\chi}_\ell
    \sim
    -\frac{c_\chi}
    {2(\ell+1)\sqrt{\Delta_\beta}}\,
    \tilde r^{\ell+1}
    \frac{e^{-\sqrt{\Delta_\beta}\tilde r}}
    {\tilde r^{(d-1)/2+\nu}} \, .
\end{equation}
Thus the particular solution dominates over the decaying homogeneous solution at sufficiently large radius in all $d$. The negative sign of the particular solution is consistent with the positive response found numerically, but we will use the numerical extraction of the asymptotic potential to determine the sign of $\lambda_\ell$.

The response potential $\delta \tilde{\varphi}_\ell$ must carry no growing tidal piece, since the growing solution already sits in the external source $\eta$. Its source $2\,\tilde{\chi}\,\delta\tilde{\chi}_\ell$ in the second equation is exponentially suppressed, so $\delta\tilde{\varphi}_\ell$ obeys the vacuum Laplace equation asymptotically and is purely decaying,
\begin{equation}
\label{eq:devphi_asymptotic}
    \delta\tilde{\varphi}_\ell(\tilde{r})
    \sim
    \frac{\mathcal{C}_\ell^\varphi}{\tilde{r}^{\ell+d-2}}
    \qquad
    \text{for } \tilde{r}\to\infty \, .
\end{equation}
The boundary conditions are thus
\begin{equation}
\label{eq:BCsPert_1}
    \delta\tilde{\chi}_\ell(\tilde{r})
    =
    a_\ell \tilde{r}^{\ell}
    +
    O(\tilde{r}^{\ell+2}) \, ,
    \qquad
    \delta\tilde{\varphi}_\ell(\tilde{r})
    =
    b_\ell \tilde{r}^{\ell}
    +
    O(\tilde{r}^{\ell+2}) \, ,
    \qquad
    \tilde{r}\to0 \, ,
\end{equation}
and
\begin{equation}
\label{eq:BCsPert_2}
    \delta\tilde{\chi}_\ell(\tilde r)
    \sim
    -\frac{c_\chi}
    {2(\ell+1)\sqrt{\Delta_\beta}}\,
    e^{-\sqrt{\Delta_\beta}\tilde r}
    \tilde r^{\ell+1-(d-1)/2-\nu} \, ,
    \qquad
    \delta\tilde{\varphi}_\ell(\tilde r)
    \sim
    \frac{\mathcal{C}_\ell^\varphi}
    {\tilde r^{\ell+d-2}} \, ,
    \qquad
    \tilde r\to\infty \, .
\end{equation}
The coefficient
\begin{equation}
\label{eq:Cell_def}
    \mathcal{C}_\ell^\varphi
    =
    \lim_{\tilde{r}\rightarrow\infty}
    \tilde{r}^{\ell+d-2}
    \, \delta\tilde{\varphi}_\ell(\tilde{r})
\end{equation}
is the asymptotic response coefficient from which we extract the Love number.

Numerically, the origin is replaced by a small radius $\tilde{r}_{\min}$ and infinity by a large radius $\tilde{r}_{\max}$. Rather than fix $a_\ell$ and $b_\ell$ in advance, we use them as shooting parameters, imposing regularity through
\begin{equation}
\label{eq:dchi_BCs_rmin}
    \delta\tilde{\chi}_\ell(\tilde{r}_{\min})
    =
    a_\ell \, \tilde{r}_{\min}^{\ell} \, ,
    \qquad
    \delta\tilde{\chi}'_\ell(\tilde{r}_{\min})
    =
    \ell a_\ell \, \tilde{r}_{\min}^{\ell-1} \, ,
\end{equation}
\begin{equation}
\label{eq:dphi_BCs_rmin}
    \delta\tilde{\varphi}_\ell(\tilde{r}_{\min})
    =
    b_\ell \, \tilde{r}_{\min}^{\ell} \, ,
    \qquad
    \delta\tilde{\varphi}'_\ell(\tilde{r}_{\min})
    =
    \ell b_\ell \, \tilde{r}_{\min}^{\ell-1} \, .
\end{equation}
The two parameters are then fixed by the large-radius conditions: decay of the condensate perturbation,\footnote{The condition \eqref{eq:dchi_BC_rmax} approximates the exact asymptotic behavior \eqref{eq:dechi_asymptotic}: since $\delta\tilde{\chi}_\ell$ is exponentially suppressed, setting it to zero at sufficiently large $\tilde r_{\max}$ is exponentially accurate. Stability under variations of $\tilde r_{\max}$ is verified in Appendix~\ref{app:numerics}.}
\begin{equation}
\label{eq:dchi_BC_rmax}
    \delta\tilde{\chi}_\ell(\tilde{r}_{\max})=0 \, ,
\end{equation}
and the finite-radius form of the decaying multipolar tail for the response potential,
\begin{equation}
\label{eq:devphi_robin}
    \tilde{r}_{\max} \,
    \delta\tilde{\varphi}'_\ell(\tilde{r}_{\max})
    +
    (\ell+d-2) \,
    \delta\tilde{\varphi}_\ell(\tilde{r}_{\max})
    =
    0 \, .
\end{equation}

\subsection{Extracting the Love numbers} To express $\lambda_\ell$ in terms of the response coefficient $\mathcal{C}_\ell^\varphi$, note that \eqref{eq:tilde_rho_U} gives the density and potential perturbations
\begin{equation}
\label{eq:deltaU_deltaphi}
    \delta\tilde{\rho}_\ell = \frac{\mathcal{M}_s^{2-d}}{4 \pi G_N} \, \tilde{\chi} \, \delta\tilde{\chi}_\ell \, , \qquad \delta\tilde{U}_\ell
    =
    \frac{\omega_{d-1} \, \mathcal{M}_s^{2-d}}{8\pi} \,
    \delta\tilde{\varphi}_\ell \, ,
\end{equation}
so that \eqref{eq:devphi_asymptotic} yields
\begin{equation}
    \delta\tilde{U}_\ell
    \sim
    \frac{\omega_{d-1} \, \mathcal{M}_s^{2-d}}{8\pi} \,
    \frac{\mathcal{C}_\ell^\varphi}{\tilde{r}^{\ell+d-2}}
    \qquad
    \text{for } \tilde{r}\to\infty \, .
\end{equation}
Comparing with \eqref{eq:deltaUinfty},
\begin{equation}
\label{eq:B_Cell_relation}
    B_\ell
    =
    -
    \frac{(2\ell+d-2)\mathcal{M}_s^{2-d}}{8\pi G_N}
    \, \mathcal{C}_\ell^\varphi \, .
\end{equation}
Combining the definition \eqref{eq:TLNsdef} with \eqref{eq:omega_unit_sphere}, \eqref{eq:rb}, \eqref{eq:fell_choice}, and \eqref{eq:deltaU_deltaphi}--\eqref{eq:B_Cell_relation} gives the formula used to determine the Love numbers numerically:
\begin{align}
    \label{eq:TLN_Cell}
    \lambda_\ell &=
    \frac{2\ell+d-2}{2 \, \omega_{d-1} \, \tilde{r}_{\text{\tiny HP}}^{2\ell+d-2}}
    \,
    \mathcal{C}_\ell^\varphi\\
    \label{eq:TLNsfinal}
    &= \frac{(2\ell+d-2) \, \Gamma(d/2)}{4\pi^{d/2}} \,
    \Delta_\beta^{\frac{2\ell+d-2}{2}} \,
    \lim_{\tilde{r}\rightarrow\infty}
    \tilde{r}^{\ell+d-2} \,
    \delta\tilde{\varphi}_\ell(\tilde{r}) \, .
\end{align}

\subsection{A deformability radius} The dimensionless Love numbers depend homogeneously on the length chosen to
normalize the response: if the reference radius is changed as
$\tilde r_{\text{\tiny HP}}\to\#\,\tilde r_{\text{\tiny HP}}$, then $\lambda_\ell\to\#^{-(2\ell+d-2)}\lambda_\ell$. Their numerical values therefore depend on this choice of scale, with the convention dependence becoming increasingly pronounced at higher multipoles. For an object with a sharp surface this is harmless, since the
normalizing length is unambiguous. Here it is not: as noted above,
$\tilde{r}_{\text{\tiny HP}}$ is a characteristic radius rather than a boundary, and any
other choice differing from it by a factor $\#$ rescales $\lambda_\ell$ by
$\#^{-(2\ell+d-2)}$. Comparisons across $\ell$ therefore mix the physical response
with a growing power of a conventionally chosen length.

This convention dependence is removed by converting the response back into a physical length. We define the deformability radius
\begin{equation}
\label{eq:Rell_def}
    \mathcal{R}_\ell
    \equiv
    r_{\text{\tiny HP}} \,
    \lambda_\ell^{1/(2\ell+d-2)}
    =
    \frac{\tilde r_{\text{\tiny HP}}}{\mathcal{M}_s} \,
    \lambda_\ell^{1/(2\ell+d-2)} \, .
\end{equation}
This is invariant under a rescaling of the reference radius used to render the response dimensionless. The ratio
$\mathcal{R}_\ell/r_{\text{\tiny HP}}$ then measures, in units of the size of the star, the radial scale associated with the $\ell$-th multipolar response.

\subsection{Numerical results} The computation proceeds in three steps. We first solve the background equations \eqref{eq:EOMUnpert} with boundary conditions \eqref{eq:BCsUnpert}, obtaining $\tilde{\chi}(\tilde{r})$, $\tilde{\varphi}(\tilde{r})$, and $\Delta_\beta=\tilde{\varphi}(\infty)$. For each fixed $\ell$ we then solve the perturbation equations \eqref{eq:EOMPertSH}, implementing the analytical boundary conditions \eqref{eq:BCsPert_1}--\eqref{eq:BCsPert_2} at the finite radii $\tilde r_{\min}$ and $\tilde r_{\max}$ as described above, cf.~\eqref{eq:dchi_BCs_rmin}--\eqref{eq:devphi_robin}. Finally we extract $\mathcal{C}_\ell^\varphi$ from the large-radius behavior of $\delta\tilde{\varphi}_\ell$ and compute $\lambda_\ell$ from \eqref{eq:TLN_Cell}--\eqref{eq:TLNsfinal}. The shooting procedure, plateau extraction, and numerical stability estimates are detailed in Appendix~\ref{app:numerics}.

\begin{figure}[t!]
    \centering
    \includegraphics[width=0.475\textwidth]{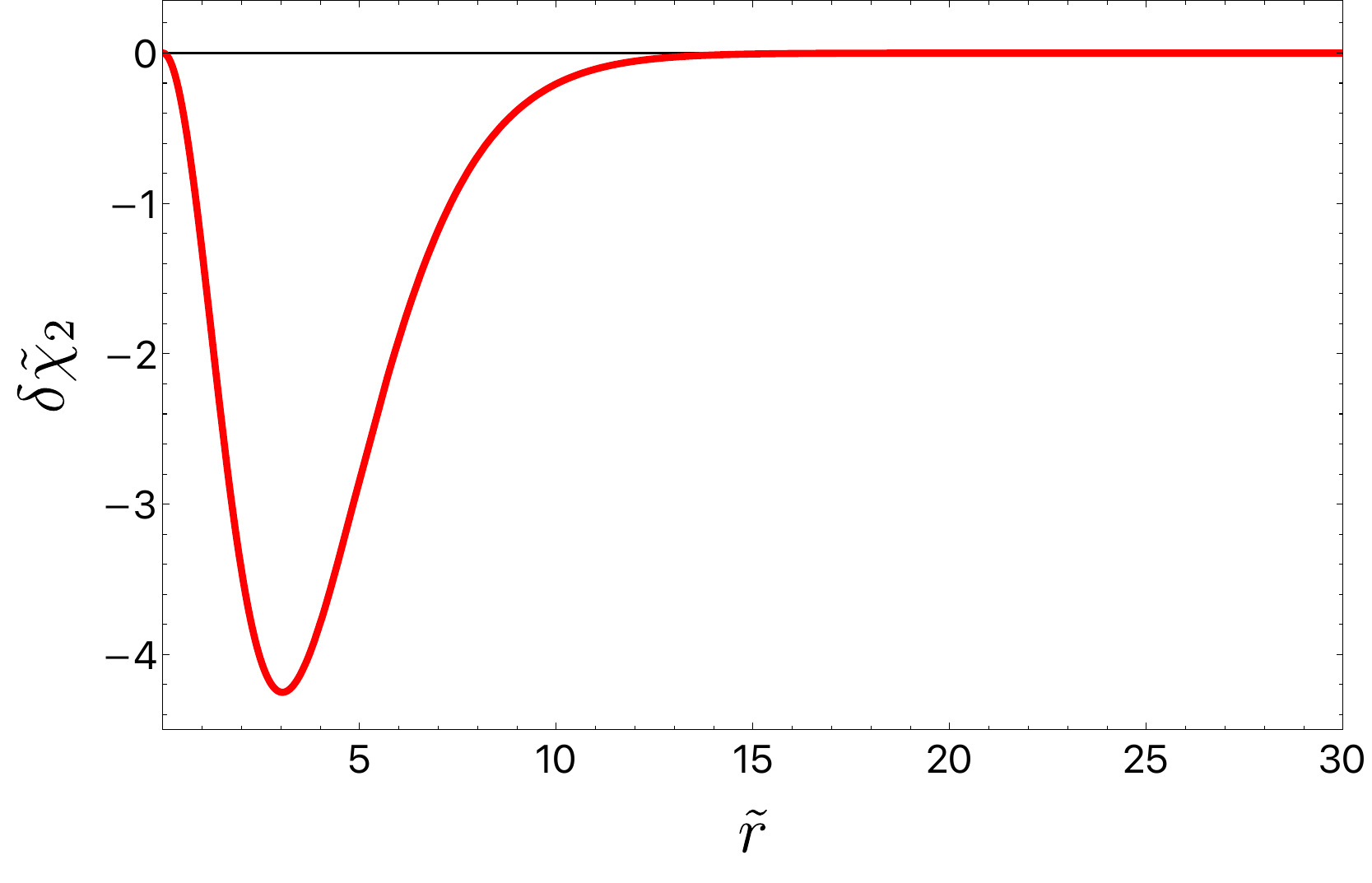}\qquad\includegraphics[width=0.475\textwidth]{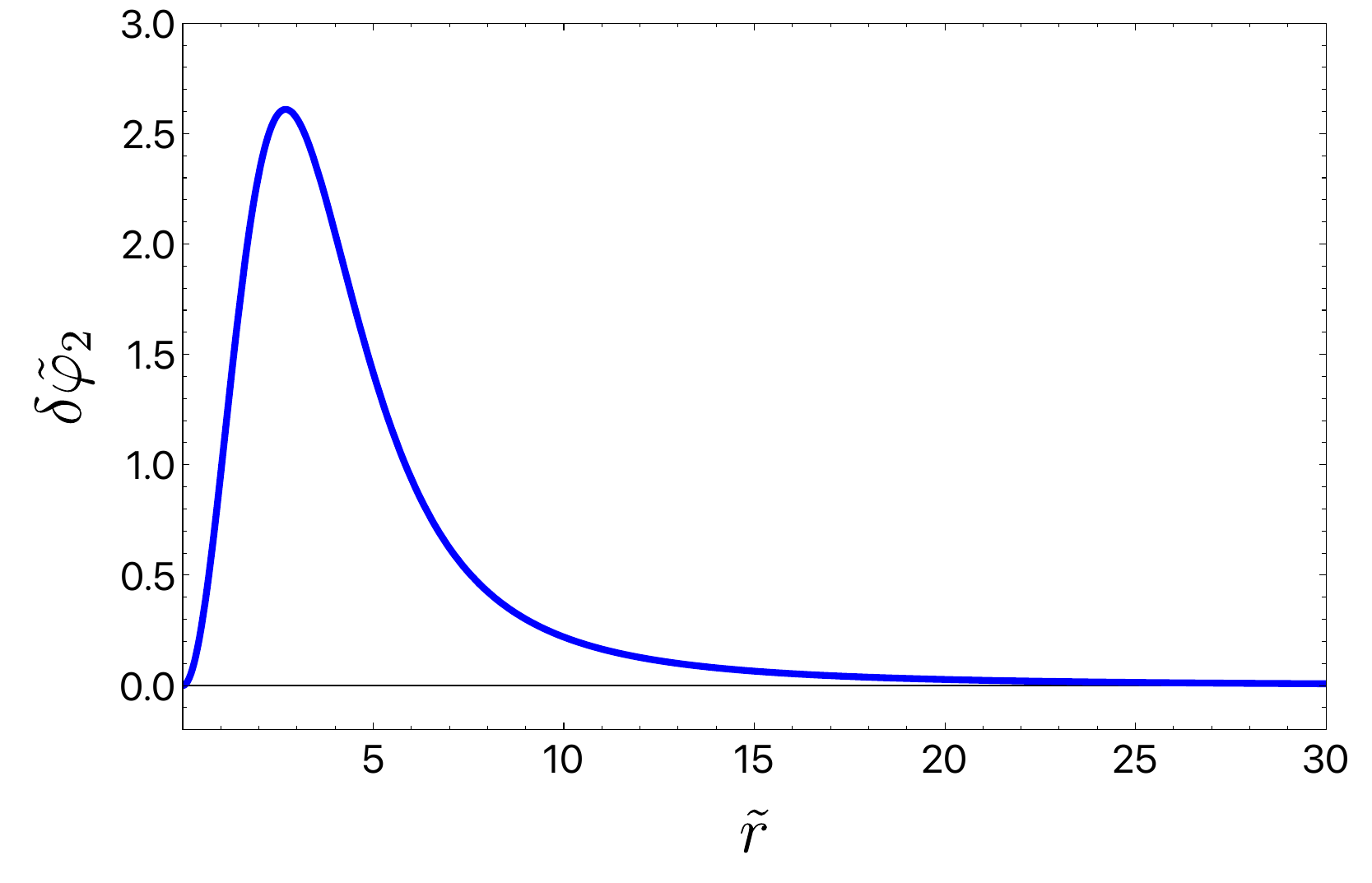}
    \caption{\small Quadrupolar perturbation of the HP string star in $D=4$. We show the condensate perturbation $\delta\tilde{\chi}_2$ (left) and response potential $\delta\tilde{\varphi}_2$ (right) as functions of $\tilde{r}$, obtained by solving \eqref{eq:EOMPertSH}. The analytical boundary conditions \eqref{eq:BCsPert_1}--\eqref{eq:BCsPert_2} are implemented at finite radii through \eqref{eq:dchi_BCs_rmin}--\eqref{eq:devphi_robin}. The applied tidal potential is \eqref{eq:Vext_choice}, and the induced deformation of the mass density and Newtonian potential is given by \eqref{eq:deltaU_deltaphi}. The rescaled variables relate to the originals $r,\chi,\varphi$ via \eqref{eq:tilde_variables_def}.}
\label{fig:delta_chi_delta_phi}
\end{figure}

Figure~\ref{fig:delta_chi_delta_phi} shows the quadrupolar perturbations $\delta\tilde{\chi}_2$ and $\delta\tilde{\varphi}_2$ in $D=4$. As expected from \eqref{eq:dechi_asymptotic} and \eqref{eq:devphi_asymptotic}, the condensate perturbation decays exponentially and the response potential algebraically.

\begin{figure}[t!]
    \centering
    \includegraphics[width=0.65\textwidth]{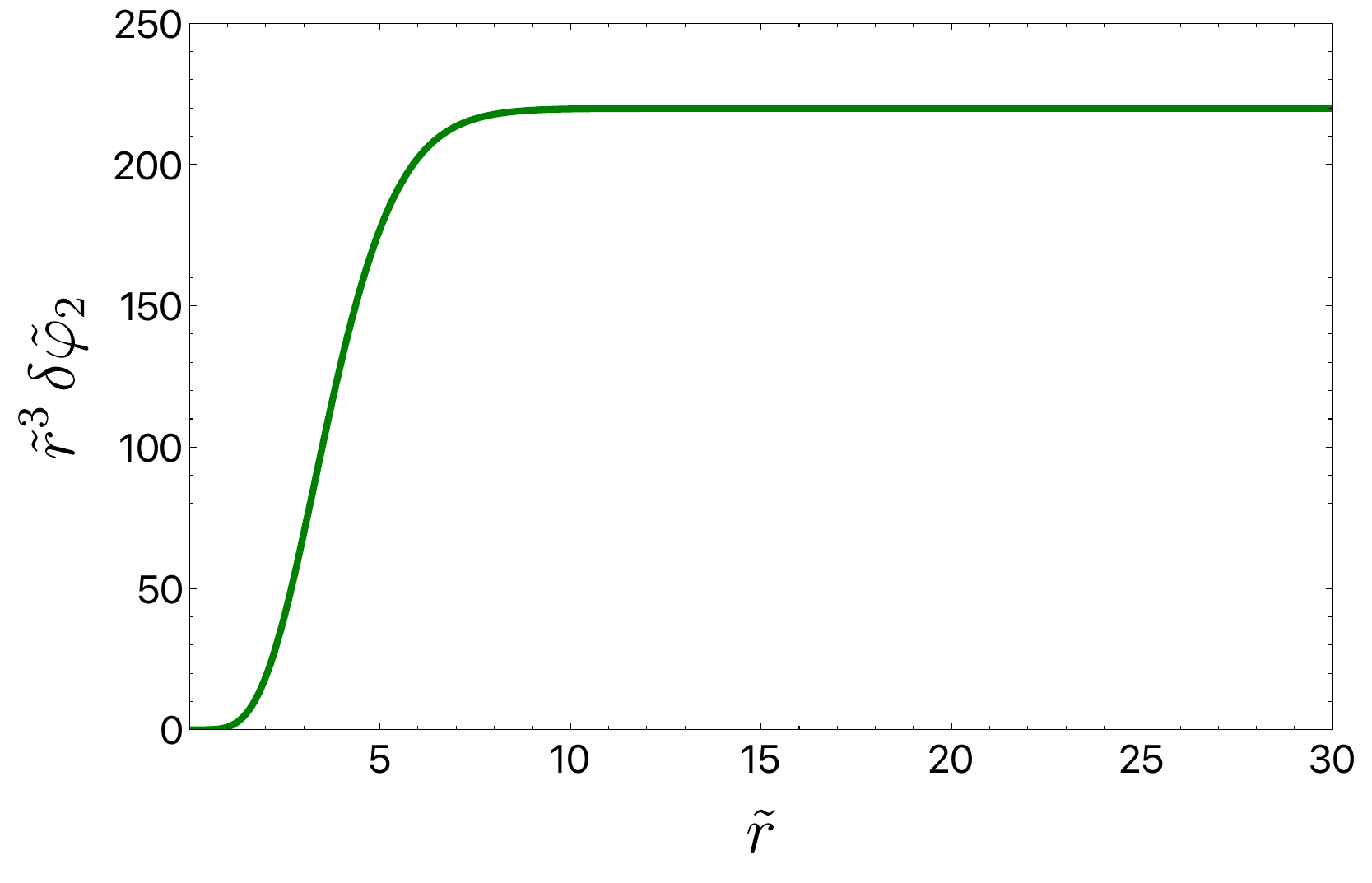}
    \caption{\small Plateau used to extract the response coefficient $\mathcal{C}_2^\varphi$ in $D=4$, cf.~\eqref{eq:Cell_def}. We plot $\tilde{r}^{3}\delta\tilde{\varphi}_2(\tilde{r})$ against $\tilde{r}$; in the exact solution this tends to a constant at large radius, equal to $\mathcal{C}_2^\varphi$. We estimate $\mathcal{C}_2^\varphi$ from the approximately constant plateau and compute the Love number from \eqref{eq:TLN_Cell}--\eqref{eq:TLNsfinal}.}
    \label{fig:plateau}
\end{figure}

In the exact solution $\tilde{r}^{\ell+d-2}\delta\tilde{\varphi}_\ell(\tilde{r})$ tends to the constant $\mathcal{C}_\ell^\varphi$ at large radius. We extract $\mathcal{C}_\ell^\varphi$ from the plateau of this quantity, in the region where the solution has reached its asymptotic decaying regime but before the finite-radius boundary condition distorts the profile; Figure~\ref{fig:plateau} gives an example for the $D=4$ quadrupole.

Carrying this out for $\ell=2,3,4$ in $D=4,5,6$ gives the Love numbers in Table~\ref{tab:HPLoveNumbers}, which admit a simple qualitative reading. First, all are non-zero—the expected behavior for an extended horizonless object: the tidal field polarizes the mass distribution and induces a non-vanishing multipolar response in the asymptotic Newtonian potential. In our sign convention this response gives positive $\lambda_\ell$.

Second, at fixed $\ell$, $\lambda_\ell$ decreases from $D=4$ to $D=6$. From \eqref{eq:TLN_Cell} this follows from the interplay of $\mathcal{C}_\ell^\varphi$ and $\tilde{r}_{\text{\tiny HP}}$ with dimension. The response coefficient $\mathcal{C}_\ell^\varphi$ grows with $D$ (Table~\ref{tab:CphiNumerics}), but so does the characteristic size $\tilde{r}_{\text{\tiny HP}}$ (Table~\ref{tab:HPBackgroundData}), and the power $2\ell+d-2$ in \eqref{eq:TLN_Cell} strongly suppresses the result; the prefactor $(2\ell+d-2)/(2\omega_{d-1})$ also decreases mildly across this range. The net effect is that the dimensionless polarizability, measured in units of $\tilde{r}_{\text{\tiny HP}}$, decreases with $D$—consistent with the Newtonian potential becoming shorter-ranged as $D$ grows.

\begin{table}[t!]
\centering
\small
\begin{tabular}{|c||c|c|c|}
    \hline
    $\lambda_\ell$ & $\ell=2$ & $\ell=3$ & $\ell=4$ \\
    \hline\hline
    $D=4$ & $41.5$ & $626$ & $1.68\times10^4$ \\
    \hline
    $D=5$ & $21.5$ & $216$ & $3.90 \times 10^3$ \\
    \hline
    $D=6$ & $6.32$ & $50.5$ & $750$ \\
    \hline
\end{tabular}
\caption{\small Static tidal Love numbers of the HP string star in $D=4,5,6$ for the multipoles $\ell=2,3,4$. The displayed values are rounded to three significant figures. Their numerical extraction is stable at a relative level of approximately $2.5\times10^{-3}$ or better under the variations described in Appendix~\ref{app:numerics}; this should be regarded as a stability diagnostic rather than as a complete estimate of the total numerical error.}
\label{tab:HPLoveNumbers}
\end{table}

Third, at fixed $D$, the Love numbers grow rapidly with $\ell$, driven mainly by the rapid increase of $\mathcal{C}_\ell^\varphi$. The origin of this behavior can be understood from the radial support of the perturbation. Since the tidal field has the form $\eta_\ell(\tilde r)=\tilde r^\ell$, the source in the first equation of \eqref{eq:EOMPertSH} carries the factor $\tilde r^\ell\tilde\chi$. Increasing $\ell$ therefore shifts outward the effective support of the forcing, until the exponential tail of the condensate suppresses it.
The same outward displacement is amplified in the induced multipole: $\mathcal{C}_\ell^\varphi$ is proportional to $B_\ell$ through
\eqref{eq:B_Cell_relation}, while and $B_\ell$ contains the additional radial weight $\tilde r^{\ell+d-1}$ in \eqref{eq:Ilm}. 

Thus the multipolar response is controlled by a competition between the power-law enhancement associated with the tidal field and the induced multipole moment, and the exponential suppression of the condensate tail. The numerical results show that, for $\ell=2,3,4$, the former wins. We now make this competition explicit analytically.

\begin{table}[t]
\centering
\begin{tabular}{|c||c|c|c||c|c|}
\hline
$D$
& $\mathcal{R}_2/r_{\text{\tiny HP}}$
& $\mathcal{R}_3/r_{\text{\tiny HP}}$
& $\mathcal{R}_4/r_{\text{\tiny HP}}$
& $\Delta\mathcal{R}_{23}/r_{\text{\tiny HP}}$
& $\Delta\mathcal{R}_{34}/r_{\text{\tiny HP}}$
\\
\hline
4 & 2.11 & 2.51 & 2.95 & 0.40 & 0.44 \\
5 & 1.67 & 1.96 & 2.29 & 0.29 & 0.33 \\
6 & 1.30 &  1.55 & 1.83 & 0.25 & 0.28 \\
\hline
\end{tabular}
\caption{\small
Deformability radius $\mathcal{R}_\ell/r_{\text{\tiny HP}}$ of the HP
string star for $\ell=2,3,4$, obtained from the numerical Love numbers
in Table~\ref{tab:HPLoveNumbers}. The last two columns give the
successive differences,
$\Delta\mathcal{R}_{23}=\mathcal{R}_3-\mathcal{R}_2$ and
$\Delta\mathcal{R}_{34}=\mathcal{R}_4-\mathcal{R}_3$, which
show the approximately
linear growth with $\ell$.
}
\label{tab:Rell}
\end{table}

\subsection{Multipolar growth from the condensate tail}
The behavior of the deformability radius introduced in \eqref{eq:Rell_def} can be understood analytically from the large-radius asymptotics of the condensate and its forced perturbation.

At fixed $\ell$ and large $\tilde r$, using \eqref{eq:chNI} and \eqref{eq:dechi_asymptotic}, the background and forced perturbation behave as
\begin{equation}
    \tilde{\chi}
    \sim
    c_\chi
    \frac{e^{-\sqrt{\Delta_\beta}\tilde r}}
    {\tilde r^{\frac{d-1}{2}+\nu}} \, ,
    \qquad
    \delta\tilde{\chi}_\ell
    \sim
    -\frac{c_\chi}{2(\ell+1)\sqrt{\Delta_\beta}}\,
    e^{-\sqrt{\Delta_\beta}\tilde r}
    \tilde r^{\ell+1-\frac{d-1}{2}-\nu}\,.
\end{equation}
The induced multipole moment \eqref{eq:Ilm} is proportional to
\begin{equation}
    B_\ell
    \propto
    \int_0^\infty d\tilde r\,
    \tilde r^{\ell+d-1}
    \tilde\chi\,\delta\tilde\chi_\ell \, .
\end{equation}
Consequently, the powers associated with the dimension cancel and, up to a $\tilde r$-independent factor, the large-radius integrand takes the universal form
\begin{equation}
\label{eq:Bell_integrand}
    \tilde r^{\ell+d-1}
    \tilde\chi\,\delta\tilde\chi_\ell
    \propto
    \tilde r^{2\ell+1-2\nu}
    e^{-2\sqrt{\Delta_\beta}\tilde r}\,.
\end{equation}
The Coulomb correction in $D=4$ modifies the power-law prefactor through $\nu$, but not the essential competition between multipolar growth and exponential decay. For any fixed $\ell$, the power-law factor enhances the contribution from larger radii, whereas the exponential tail eventually suppresses it, so that their competition selects a characteristic radius.

The maximum of the asymptotic profile \eqref{eq:Bell_integrand} occurs at
\begin{equation}
\label{eq:saddle_fixed_ell}
    \tilde r_*
    =
    \frac{2\ell+1-2\nu}{2\sqrt{\Delta_\beta}}
    =
    \left(
        \ell+\frac12-\nu
    \right)
    \tilde r_{\text{\tiny HP}} \, .
\end{equation}
This fixed-$\ell$ tail estimate indicates that higher multipoles probe progressively farther into the condensate and suggests the scaling
$\tilde r_*\sim\ell\,\tilde r_{\text{\tiny HP}}$.

Strictly speaking, the fixed-$\ell$ large-radius expansion is not uniform in $\ell$. While at fixed $\ell$ the centrifugal term in \eqref{eq:approx_eq_delta_chi} can be neglected at large radius, for radii of order $\tilde r\sim\ell\,\tilde r_{\text{\tiny HP}}$ it must be
retained since it is of the same order as $\Delta_\beta$. For a controlled large-$\ell$ statement, we therefore return to \eqref{eq:approx_eq_delta_chi} and perform a uniform expansion with
\begin{equation}
    \ell\to\infty \, ,
    \qquad
    \tilde r\to\infty \, ,
    \qquad
    \frac{\tilde r}{\ell}
    \text{ fixed} \, .
\end{equation}
Writing $\tilde r=\ell\,z$ and expanding \eqref{eq:approx_eq_delta_chi} at fixed $z$, the centrifugal term must be retained together with $\Delta_\beta$ at leading order. The leading behavior follows particularly simply from the cancellation between the centrifugal term and the terms generated by differentiating $\tilde r^\ell$. Using the asymptotic background equation for $\tilde\chi$, one finds
\begin{equation}
    \delta\tilde\chi_\ell
    \simeq
    -\frac{1}{2\ell\sqrt{\Delta_\beta}}\,
    \tilde r^{\ell+1}\tilde\chi
    \left[1+O\!\left(\frac{1}{\ell}\right)\right] .
\end{equation}
Thus the uniform asymptotic behavior differs from the fixed-$\ell$ tail only by relative corrections of order $1/\ell$ and yields the same leading saddle \eqref{eq:saddle_fixed_ell}, while allowing us to quantify the remainder and the large-$\ell$ scaling in a controlled way as
\begin{equation}
\label{eq:saddle}
    \tilde r_*
    =
    \left(
        \ell+\frac12-\nu
    \right)
    \tilde r_{\text{\tiny HP}}
    +
    O\!\left(
        \frac{\tilde r_{\text{\tiny HP}}}{\ell}
    \right) \, ,
    \qquad
    \frac{r_*}{r_{\text{\tiny HP}}}
    =
    \ell+O(1)\,.
\end{equation}
Thus the multipolar enhancement pushes the response outward with increasing $\ell$, while the exponential condensate tail ultimately suppresses it, and the balance between these two effects occurs at radii growing linearly with the multipole order.

Performing the asymptotic integral then gives, up to $\ell$-independent factors,
\begin{equation}
\label{eq:Bell_asymptotic}
    B_\ell
    \propto
    \frac{\Gamma(2\ell+2-2\nu)}{\ell}
    \left(
        \frac{\tilde r_{\text{\tiny HP}}}{2}
    \right)^{2\ell+2-2\nu}
    \left[
        1+O\!\left(\ell^{-1}\right)
    \right] \, .
\end{equation}
The factorial growth in this expression explains the rapid increase of the Love numbers with $\ell$. Since the normalization of the applied field is independent of $\ell$, \eqref{eq:TLNsdef} implies
$\lambda_\ell\propto B_\ell/\tilde r_{\text{\tiny HP}}^{2\ell+d-2}$, and Stirling's expansion therefore gives
\begin{equation}
\label{eq:lnlambda}
    \ln\lambda_\ell
    =
    2\ell\ln\ell
    -
    2\ell
    +
    O(\ln\ell)\,.
\end{equation}
The deformability radius consequently satisfies
\begin{equation}
\label{eq:Rell_asymptotic}
    \frac{\mathcal R_\ell}
    {r_{\text{\tiny HP}}}
    =
    \frac{\ell}{e}
    +
    O(\ln\ell) \, .
\end{equation}
The coefficient $1/e$ is universal at leading order. The Coulomb correction in $D=4$, as well as other dimension-dependent prefactors and subleading terms, affects the subleading behavior and the approach to the asymptotic regime but not the leading linear slope.

The comparison with the numerical results is best made through the successive differences of $\mathcal R_\ell$, which provide a measure of the local slope. The values are collected in Table~\ref{tab:Rell}. The successive increments are $0.40$ and $0.44$ in $D=4$, $0.29$ and $0.33$ in $D=5$, and $0.25$ and $0.28$ in $D=6$, showing an approximately linear increase over the multipoles studied. The asymptotic analysis predicts a linear increase with the universal large-$\ell$ slope $1/e\simeq0.368$. The available data are compatible with this behavior, but with only $\ell=2,3,4$ they do not yet provide a precise quantitative test of the asymptotic slope, since dimension-dependent subleading terms can still be important at these low multipoles.

\section{Tidal response across the black hole--string transition}
\label{sec:BHstringComparison}

The HP string star describes at weak coupling the same states that appear at stronger coupling as a static neutral black hole, so it is natural to compare the HP Love numbers of Section~\ref{sec:TLNsHP} with known results for the response of static neutral black holes. The comparison is necessarily qualitative, since neither the HP effective theory nor the perturbative $\alpha'$ expansion are quantitatively controlled near the correspondence point, where the characteristic size of the configuration becomes string-scale,
\begin{equation}
\label{eq:corr_pt}
    r_{\text{\tiny HP}}\sim r_{\text{\tiny BH}} \sim \ell_s = \sqrt{\alpha'} \,.
\end{equation}
On the black hole side, higher-curvature $\alpha'$ corrections become unsuppressed. On the string side, the separation between the light winding mode retained in the HP truncation and the other string modes disappears when $\Delta_\beta =O(1)$, and higher-derivative and higher-order interactions can no longer be neglected. In both descriptions the separation of scales underlying the effective theory has closed, so a direct numerical matching cannot be expected. The question we can ask is nonetheless well posed: what becomes of the fine-tuned vanishing of the black hole tidal response as one moves toward the stringy regime?

On the string side, $\lambda_\ell \equiv \lambda_\ell^{\text{\tiny HP}}$ is the response of the Newtonian potential of the star to an external tidal potential, defined within the leading HP description; on the black hole side, the deformability is captured by relativistic tidal Love numbers. Since we compare only qualitative features---zero versus non-zero response, and the behavior with $\ell$---any overall normalization independent of $\ell$ drops out of the comparison.

\subsection{Pure Einstein gravity} Recall first the Schwarzschild result in $D=4$. In pure Einstein gravity the static tidal Love numbers all vanish,
\begin{equation}
\label{eq:Schwarzschild_zero_Love}
    \lambda_\ell^{\text{\tiny GR}}=0
    \qquad
    (D=4, \, \ell\geq2) \, ,
\end{equation}
in the standard relativistic definition of the scalar-type (even parity, electric-type) and vector-type (odd parity, magnetic-type) Love numbers \cite{Binnington:2009bb,Hui:2020xxx}. For the four-dimensional HP string star, by contrast, we found the non-zero values in \eqref{eq:TLN_D4_intro}.
The transition from Schwarzschild to the HP string star thus changes the static response from zero to non-zero: in this precise sense, the exceptional tidal rigidity of the four-dimensional Schwarzschild solution is lost on the string side.

In higher dimensions the black hole response already has richer structure. By the Kodama-Ishibashi classification, a tidal perturbation in $D \ge 5$ acquires a tensor-type sector absent in $D=4$, where only scalar- and vector-type sectors appear \cite{Kodama:2003jz,Ishibashi:2011ws}. Moreover, the Love numbers of the Schwarzschild-Tangherlini black hole in $D \ge 5$ can be non-zero, and can run with a renormalization scale already classically \cite{Kol:2011vg,Hui:2020xxx}.

The HP Love numbers, computed in the Newtonian description, should be compared with the relativistic channel that reduces, in the appropriate limit, to the response of mass multipoles to an even-parity tidal field—the scalar-type sector in the Kodama-Ishibashi classification. The vector-type sector governs the current-type response, with no direct Newtonian analogue, and tensor-type perturbations exist only for $D \ge 5$ as an intrinsically higher-dimensional channel. Both should therefore be regarded as purely relativistic; the natural relativistic continuation of the Newtonian mass-multipole response is the scalar-type sector.\footnote{Strictly, the HP radion response uplifts to a fixed graviton--dilaton combination, as mentioned in footnote~\ref{radion_caveat}. A like-for-like relativistic comparison would require the corresponding coupled response of the string-corrected black hole.
}

Throughout this section, all black hole Love numbers are expressed in our
conventions.\footnote{\label{BH_conventions}The Love numbers in
\eqref{eq:TLN_BH_Stype}, \eqref{eq:TLN_typeII_l23}--\eqref{eq:TLN_typeII_l4} and \eqref{eq:heterotic_even_TLNs} are related to the conventions used in the
literature, respectively, by
\begin{align*}
    \lambda_\ell^{\text{\tiny GR}} &= \frac{\ell (\ell - 1) (2 \ell + D - 3)}{2 \omega_{D-2} (\ell + D - 3) (\ell + D - 2)} \, k_\text{Z} \, ,
    \qquad &&\text{\cite{Hui:2020xxx}}\\
    \lambda_\ell^{\text{\tiny BH}} &= -\frac{2 \ell + 1}{2^{2 \ell + 3} \pi} k_\ell^+ \, ,
    &&\text{\cite{Cano:2025zyk}}\\
    \lambda_\ell^{\text{\tiny BH}} &= -\frac{2 \ell + 1}{4 \pi} \kappa_\ell^+ \, .
    &&\text{\cite{Katagiri:2024fpn}}
\end{align*}
For \cite{Katagiri:2024fpn}, we also use
$\alpha_{\rm dGB}=\alpha'/8$ and $r_{\rm BH}=2M$, so that
$\zeta_{\rm dGB}\equiv\alpha_{\rm dGB}/M^2=\epsilon/2$.} The scalar-type Love numbers of the Schwarzschild-Tangherlini black hole are given by \cite{Kol:2011vg,Hui:2020xxx}
\begin{equation}
\label{eq:TLN_BH_Stype}
    \lambda_\ell^{\text{\tiny GR}} = -\frac{(2 \ell + D-3) \, (\ell + D - 2)}{2^{4 \hat{\ell} + 3} \, (\ell - 1) \, \omega_{D-2}} \, \frac{\Gamma(\hat{\ell}) \, \Gamma(\hat{\ell} + 2)}{\Gamma\!\left(\hat{\ell} + \frac{1}{2} \right) \, \Gamma\!\left(\hat{\ell} + \frac{3}{2} \right)} \, \tan(\pi \hat{\ell}) \, , \qquad \hat{\ell}=\frac{\ell}{D-3} \, .
\end{equation}
Integer $\hat{\ell}$ gives a vanishing response; half-integer $\hat{\ell}$ produces a pole in \eqref{eq:TLN_BH_Stype}, signaling a logarithm in the asymptotic solution and a Love number that runs; all other values give a finite non-zero response. The resulting pattern is dimension-dependent. In $D=4$, $\hat{\ell}=\ell$, so the scalar-type response vanishes for all integer $\ell$. In $D=5$, $\ell=2,4$ give integer $\hat{\ell}$ and vanish, while $\ell=3$ runs logarithmically. In $D=6$, $\ell=2,4$ give non-zero non-running coefficients, while $\ell=3$ vanishes.

This is very different from the HP result. In the range studied, $D=4,5,6$ and $\ell=2,3,4$, the HP Love numbers are positive and non-zero throughout: the string star response inherits neither the special zeros nor the logarithmic running of the Schwarzschild-Tangherlini black hole.

\subsection{Stringy corrections on the black hole side} Away from pure Einstein gravity, the black hole solution and its tidal response receive stringy higher-derivative corrections organized in powers of
\begin{equation}
    \epsilon \equiv \frac{\alpha'}{r_{\text{\tiny BH}}^2} \, .
\end{equation}
The relevant benchmark for the HP Love numbers is therefore the scalar-type response of the $\alpha'$-corrected solution, rather than the Einstein one. Schematically,
\begin{equation}
\label{eq:BH_alpha_expansion}
    \lambda_{\ell}^{\text{\tiny BH}} = \lambda_{\ell}^{\text{\tiny GR}} + \lambda_{\ell}^{(1)} \, \epsilon + \lambda_{\ell}^{(2)} \, \epsilon^2 + \lambda_{\ell}^{(3)} \, \epsilon^3 + O(\epsilon^4) \, ,
\end{equation}
where some of the coefficients may vanish, depending on the theory and spacetime dimension being considered. For a general EFT analysis of constant and running black hole Love numbers induced by higher-derivative operators, see also \cite{Barbosa:2025uau}.

In type II superstring theory the leading correction occurs at
$O(\epsilon^3)$ \cite{Myers:1987qx,Chen:2021qrz}. In heterotic superstring
theory, by contrast, the effective action contains higher-derivative
interactions already at $O(\epsilon)$
\cite{Callan:1988hs,Moura:2006pz,Charalambous:2024tdj}. For a neutral black
hole in four dimensions, however, the $O(\epsilon)$ terms source scalar hair, while the Einstein-frame metric and its even-parity gravitational response first change at $O(\epsilon^2)$ \cite{Cano:2021rey}.

The expansion is perturbative for large black holes, $r_{\text{\tiny BH}}\gg\ell_s$, and fails at the correspondence point for the reasons given above. The leading correction therefore cannot reproduce the HP Love numbers quantitatively; its role is diagnostic, probing whether stringy effects move the black hole response away from its vanishing Einstein value.

The scalar-type tidal Love numbers of the $\alpha'$-corrected Schwarzschild geometry have been computed in type II superstring theory in four spacetime dimensions \cite{Cano:2025zyk}; see also \cite{Cano:2022wwo,Wang:2026qst,Cano:2026hlv}. For $\ell = 2,3$, they are finite and non-zero, and are given in our conventions by (see footnote~\ref{BH_conventions})
\begin{equation}
\label{eq:TLN_typeII_l23}
    \lambda_2^{\text{\tiny BH}} = -\frac{63}{160 \pi} \, \zeta(3) \, \epsilon^3 + \cdots \, , \qquad \lambda_3^{\text{\tiny BH}} = -\frac{693}{128 \pi} \, \zeta(3) \, \epsilon^3 + \cdots \qquad (D = 4, \text{ type II}) \, ,
\end{equation}
where $\zeta(3) \approx 1.20206$ and the dots denote subleading corrections in $\epsilon \ll 1$. At the order displayed, we use $r_{\text{\tiny BH}} = 2 M$. For $\ell \ge 4$ the Love numbers run logarithmically, with beta functions given in \cite{Cano:2025zyk}. In particular, for $\ell = 4$ we have
\begin{equation}
\label{eq:TLN_typeII_l4}
    \lambda_4^{\text{\tiny BH}}(r) = -\frac{2835}{16 \pi} \, \zeta(3) \, \epsilon^3 \, \ln\!\left(\frac{r_0}{r} \right) + \cdots \qquad (D = 4, \text{ type II}) \, ,
\end{equation}
where $r_0$ is a renormalization scale to be fixed, e.g., by experimental or observational data.

There is also a four-dimensional heterotic benchmark. At first order in $\alpha'$, toroidal compactification of heterotic string theory yields an axidilaton theory in which the dilaton couples to the Gauss-Bonnet density and the axion to the Pontryagin density \cite{Cano:2021rey}. For static, spherically symmetric black holes the Pontryagin density vanishes, so the background lies entirely within the Einstein--dilaton--Gauss--Bonnet (EdGB) sector. At the perturbative level, however, the axion couples to the odd-parity sector. Consequently, the even-parity (electric-type) Love numbers computed in EdGB can be identified with those of the heterotic theory, whereas the odd-parity (magnetic-type) response requires the axion and will not be considered here. Translating the even-parity EdGB tidal response of \cite{Katagiri:2024fpn} to our conventions (see footnote~\ref{BH_conventions}) then gives
\begin{equation}
\label{eq:heterotic_even_TLNs}
    \lambda_2^{\text{\tiny BH}} \simeq 0.075 \, \epsilon^2 + \cdots \, , \qquad \lambda_3^{\text{\tiny BH}} \simeq 0.0060 \, \epsilon^2 + \cdots \qquad (D = 4, \text{ heterotic}) \, .
\end{equation}
The numerical coefficients in \eqref{eq:heterotic_even_TLNs} are quoted with
the numerical precision of \cite{Katagiri:2024fpn}. We are not aware of an
analogous $\ell=4$ heterotic result in the literature.

Observe that the sign of the response is not a robust feature on the black hole side: it is negative in type II \eqref{eq:TLN_typeII_l23}, and positive in heterotic \eqref{eq:heterotic_even_TLNs}, and already in pure Einstein gravity it alternates with $\ell$, as \eqref{eq:TLN_BH_Stype} makes clear. The HP response, by contrast, is positive in all the cases we compute, as for any polarizable mass distribution, and we accordingly use $|\lambda_\ell^{\text{\tiny BH}}|$ below.

In $D=4$, we define the black hole deformability radius by
\begin{equation}
    \mathcal{R}_\ell^{\text{\tiny BH}} = r_{\text{\tiny BH}} \, \left|\lambda_\ell^{\text{\tiny BH}} \right|^{1/(2 \ell + 1)} \, ,
\end{equation}
given explicitly by
\begin{equation}
    \frac{\mathcal{R}_2^{\text{\tiny BH}}}{r_{\text{\tiny BH}}} \simeq 0.68 \, \epsilon^{3/5} \, , \qquad \frac{\mathcal{R}_3^{\text{\tiny BH}}}{r_{\text{\tiny BH}}} \simeq 1.1 \, \epsilon^{3/7} \qquad (D = 4, \text{ type II}) \, ,
\end{equation}
\begin{equation}
    \frac{\mathcal{R}_2^{\text{\tiny BH}}}{r_{\text{\tiny BH}}} \simeq 0.60 \, \epsilon^{2/5} \, , \qquad \frac{\mathcal{R}_3^{\text{\tiny BH}}}{r_{\text{\tiny BH}}} \simeq 0.48 \, \epsilon^{2/7} \qquad (D = 4, \text{ heterotic}) \, .
\end{equation}
For sufficiently small $\epsilon$, the displayed deformability radii increase from $\ell=2$ to $\ell=3$ in both examples. This does not by itself establish a dynamical growth mechanism: part of the increase follows simply from taking the $(2\ell+1)$-th root of a fixed small power of $\epsilon$, and the available results cover too few non-running multipoles to determine a trend. There may nevertheless be a meaningful competition in the black hole perturbation problem. At radius $r$, an $\ell$-pole has angular gradients of order $\ell/r$, which can enhance higher-derivative corrections, while the curvature factors sourcing those corrections decrease away from the black hole. The resulting response depends on the balance of these effects in the radial problem and may grow, decrease, or run with $\ell$.

For comparison, consider an incompressible Newtonian fluid sphere of radius $R$. In our conventions,\footnote{The convention of \cite{YipLeung2017} is related to ours by
$\lambda_\ell = \frac{2\ell+1}{4\pi} \, k_\ell$.} its Love numbers and deformability radii are given by \cite{YipLeung2017}
\begin{equation}
\label{eq:TLN_incompr_star}
    \lambda_\ell
    =
    \frac{3(2\ell+1)}{16\pi(\ell-1)} \, ,
    \qquad
    \mathcal{R}_\ell
    =
    R\,\lambda_\ell^{1/(2\ell+1)} \, ,
\end{equation}
and hence
\begin{equation}
    \frac{\mathcal{R}_2}{R}
    \simeq
    0.78517 \, ,
    \qquad
    \frac{\mathcal{R}_3}{R}
    \simeq
    0.79955 \, ,
    \qquad
    \frac{\mathcal{R}_4}{R}
    \simeq
    0.82603 \, .
\end{equation}
At large $\ell$, $\lambda_\ell\to3/(8\pi)$ and
$\mathcal{R}_\ell/R\to1$ from below. Thus the deformability radius of
an incompressible star also increases with $\ell$, but it approaches
and saturates at the physical surface $R$. The mere increase of
$\mathcal{R}_\ell$ is therefore not unique to the string star. What
distinguishes the HP behavior is its unbounded outward migration
through a diffuse tail, with no finite surface at which the response
scale must saturate.

\subsection{Interpretation}

The comparison suggests two lessons. First, both sides of the black hole--string transition are compatible with the expectation that the zero-Love property of the four-dimensional Schwarzschild black hole does not persist once one moves away from pure Einstein gravity toward stringy physics. On the black hole side, perturbative $\alpha'$ corrections generate non-zero or running scalar-type response coefficients. On the string side, the HP string star has non-zero Love numbers already at leading order in its effective description.

Second, the HP string star has a particularly clear physical mechanism behind its multipolar dependence. Higher multipoles increasingly probe the dilute outer region of the winding condensate. As shown in
Section~\ref{sec:TLNsHP}, the induced multipole is controlled by a competition between powers of the radius and the exponential condensate tail, with a dominant radius that grows linearly at large $\ell$,
\begin{equation}
    r_* \sim \ell\,r_{\text{\tiny HP}}
\end{equation}
(see \eqref{eq:saddle}). This produces the factorial growth of the HP Love numbers and the corresponding linear growth of the deformability radius.

The available perturbative $\alpha'$-corrected scalar-type results relevant to the comparison above do not yet establish a universal large-$\ell$ trend: only a few multipoles have been computed, and they display no clear monotonic pattern. Large $\ell$, however, introduces a new parametric scale on the black hole side. An $\ell$-pole has angular gradients of order $\ell/r_{\text{\tiny BH}}$, so the higher-derivative expansion of the tidal response involves an additional parameter $\ell^2\alpha'/r_{\text{\tiny BH}}^2=\ell^2\epsilon$. Consequently, at fixed order in the $\alpha'$ expansion the large-$\ell$ limit is expected to become non-uniform when
\begin{equation}
    r_{\text{\tiny BH}} \sim \ell\sqrt{\alpha'} \, ,
\end{equation} 
which is parametrically before the black hole background itself reaches the usual correspondence point for $\ell\gg1$. In this sense, the onset of stringy effects in tidal observables is not uniform in multipole number.

Intriguingly, the HP result contains the same parametric scale from the string side. Extrapolating the HP scaling to the correspondence point, $r_{\text{\tiny HP}}\sim\sqrt{\alpha'}$, the dominant response radius becomes $r_*\sim\ell\sqrt{\alpha'}$. Equivalently, the angular wavelength of the $\ell$-pole at the radius dominating the HP response satisfies $r_*/\ell\sim r_{\text{\tiny HP}}\sim\sqrt{\alpha'}$. Whether this coincidence reflects a deeper matching of the large-$\ell$ tidal response across the transition, or simply the common angular scaling of multipolar perturbations, remains to be understood.

\subsection{String stars vs.\ microstate geometries} Finally, it is natural to ask how this compares with microstate geometries, a different class of horizonless configurations in string theory meant as descriptions of black hole states \cite{Bena:2022rna}. Their response to external perturbations has been studied in several cases, with outcomes that differ from one geometry to another: non-zero static Love numbers for two-charge fuzzballs \cite{Bianchi:2022qph}, vanishing static but non-trivial dynamical Love numbers for topological stars \cite{Bianchi:2023sfs}, and resonant
response associated with metastable bound states \cite{DiRusso:2024hmd}. However, any comparison between the properties of these geometries and of string stars should be made with care, since the fields and response channels being probed are not identical in all cases. More importantly, these geometries do not provide the same continuation of a neutral black hole into the stringy regime: the 2-charge geometries are supersymmetric, and describe states whose black hole counterpart has no macroscopic horizon, while top stars are macroscopic horizonless objects meant to replace a charged black hole even when this is much larger than the string scale. These configurations therefore probe a different question from the one asked here. The HP string star is distinguished by the fact that it is the weak-coupling description of the same family of neutral black hole states near the correspondence point.

\section{Conclusions and outlook}
\label{sec:conclusions}

We have computed the static tidal Love numbers of the Horowitz--Polchinski string star, which is the weak-coupling string-side configuration expected to replace a static neutral black hole across the black hole--string transition. Within the leading near-Hagedorn effective theory,\footnote{It would be interesting to reproduce the tidal response in the approach of \cite{Damour:1999aw} to self-gravitating string states.} we solved the linear response of the winding condensate and of the potential it generates to an external multipolar tidal field, for $\ell=2,3,4$ in
$D=4,5,6$ (Table~\ref{tab:HPLoveNumbers}).

The Love numbers are non-zero in every case. The tidal rigidity of the four-dimensional Schwarzschild black hole therefore does not survive the transition: once the horizon is replaced by an extended winding condensate, the system polarizes. The available $\alpha'$ corrections on the black hole side point in the same direction, so both descriptions, each within its own regime of applicability, indicate that the zero-response structure of Einstein gravity does not persist as one moves toward the stringy regime.

The most distinctive feature of the string star response is its multipolar structure. The Love numbers grow rapidly with $\ell$, and we have shown analytically that this behavior reflects a systematic outward migration of the tidal response through the winding condensate. The competition between the $r^\ell$ growth of the tidal field and the exponential fall-off of the condensate localizes the dominant contribution to the induced multipole at radii
\begin{equation}
     r_*\sim\ell\, r_{\text{\tiny HP}}\,,
\end{equation}
so that the associated deformability radius grows linearly with multipole number, with asymptotic slope $1/e$, eq.~\eqref{eq:Rell_asymptotic}. The numerical results for $\ell=2,3,4$ are consistent with this picture (Table~\ref{tab:Rell}).

This gives the tidal response a direct geometric interpretation: higher multipoles probe progressively farther into the dilute outer region of the string star. Stringy corrections on the black hole side also lift the special zero-Love property of four-dimensional Einstein gravity, but the presently available perturbative results do not yet reveal whether their multipolar response is organized in an analogous way. The relevant comparison is therefore not simply whether the black hole Love numbers grow with $\ell$, but whether such growth, if present, has a similarly direct geometric origin in the radial structure of the response.

At large multipole number, the correspondence of tidal observables is also expected to become non-uniform in $\ell$. On the black hole side, the perturbative $\alpha'$ expansion of the tidal response becomes unreliable already when $r_{\text{\tiny BH}}\sim \ell\sqrt{\alpha'}$, parametrically before the background reaches the usual correspondence point. Extrapolating the HP scaling to the correspondence point, on the string side the dominant HP response occurs at $r_*\sim \ell\, r_{\text{\tiny HP}}\sim \ell\sqrt{\alpha'}$. The appearance of the same parametric scale in the two descriptions is intriguing and deserves further investigation.

Several extensions suggest themselves. The most immediate is rotation. The correspondence has been developed for rotating black holes and strings \cite{Ceplak:2023afb,Ceplak:2024dxm}, and self-gravitating spinning string condensates have since been constructed \cite{Santos:2024ycg,Seitz:2025wpc}. Since the Kerr Love numbers also vanish in four dimensions, and since spin opens response channels with no static counterpart, the rotating problem is a more intricate version of the question asked here. It would also be interesting to extend the present calculation to larger multipoles, both to test the asymptotic behavior of the deformability radius and to determine how the numerical response approaches the predicted large-$\ell$ regime. A fuller relativistic treatment of the coupled graviton--dilaton response would allow the two channels to be disentangled and compared directly with string-corrected black holes. Finally, it would be interesting to investigate whether the large-$\ell$ growth persists in $D\geq7$, where string star solutions exist once higher-order corrections to the leading HP effective theory are included \cite{Balthazar:2022hno,Bedroya:2024igb}. 

A further question we have not addressed is dissipation. The HP solution is horizonless and static, so it has no analogue of horizon absorption, unlike a black hole. Whether the effective description admits a meaningful dissipative response, and how it would behave across the transition, seems worth understanding: it would test the horizon interpretation from the opposite side, through the channel that the conservative Love numbers do not probe.

\section*{Acknowledgments}

We thank Pablo Cano, Davide Cassani, Yiming Chen, Juan Maldacena, and Alejandro Ruipérez for useful conversations.
RE and ST were supported by MICINN grant PID2022-136224NB-C22, and grant CEX2024-001451-M funded by MICIU/AEI/10.13039/501100011033. The project that gave rise to these results received the support of a fellowship from ``la Caixa'' Foundation (ID 100010434), awarded to ST, with code LCF/BQ/DI24/12070002. The work of LL is supported by the Natural Sciences and Engineering Research Council (NSERC) of Canada, by the Simons Foundation through Award SFI-MPS-BH-00012593-12, by the Carlo Fidani Rainer Weiss Chair at Perimeter Institute and CIFAR. 
This research was supported
in part by Perimeter Institute for Theoretical Physics.
Research at Perimeter Institute is supported in part by
the Government of Canada through the Department of
Innovation, Science and Economic Development and by
the Province of Ontario through the Ministry of Colleges
and Universities.


\appendix

\section{Numerical method}
\label{app:numerics}

This appendix details the numerical method behind Table~\ref{tab:HPLoveNumbers}.

\subsection{Background solution}

The unperturbed problem is the boundary value problem \eqref{eq:EOMUnpert}, \eqref{eq:BCsUnpert}, which we solve by shooting on $\tilde{\varphi}_0=\tilde{\varphi}(0)$. Because of the spherical-coordinate singularity at $\tilde{r}=0$, integration starts at a small finite radius, in all cases $\tilde{r}_{\min}=10^{-3}$, and runs to $\tilde{r}_{\max}^\text{unpert} = 60$, with initial data from the regular expansion \eqref{eq:chi_phi_small_r}. We tune $\tilde{\varphi}_0$ so that $\tilde{\chi}$ decays at large radius, selecting the nodeless ground state; the resulting values are in Table~\ref{tab:HPBackgroundData}.

The near-Hagedorn parameter $\Delta_\beta$ follows from the large-radius behavior of $\tilde{\varphi}$. As discussed around \eqref{eq:vpNI}--\eqref{eq:vpinf},
\begin{equation}
    \tilde{\varphi}(\tilde{r})
    \sim
    \Delta_\beta
    +
    \frac{c_\varphi}{\tilde{r}^{d-2}} \, , \qquad c_\varphi
    =
    - \frac{1}{d-2}
    \lim_{\tilde{r}\rightarrow\infty}
    \tilde{r}^{d-1}\tilde{\varphi}'(\tilde{r}) \, .
\end{equation}
At each radius $\tilde{r}_i$ in an asymptotic sampling window we subtract the leading power-law tail to obtain a local estimate,
\begin{equation}
    \Delta_\beta(\tilde{r}_i)
    =
    \tilde{\varphi}(\tilde{r}_i)
    +
    \frac{\tilde{r}_i \, \tilde{\varphi}'(\tilde{r}_i)}{d-2} \, .
\end{equation}
The value quoted in Table~\ref{tab:HPBackgroundData} is the mean over the $N$ sampled radii in the window,
\begin{equation}
\label{eq:delta_beta_best}
    \Delta_\beta
    =
    \frac{1}{N} \,
    \sum_i
    \Delta_\beta(\tilde{r}_i) \, ,
\end{equation}
with the stability of the extraction quantified by the maximum deviation across the window,
\begin{equation}
    \delta\Delta_\beta
    =
    \max_i
    \left|
    \Delta_\beta(\tilde{r}_i)
    -
    \Delta_\beta
    \right| \, .
\end{equation}
For the three dimensions $D=4,5,6$ considered here, the relative stability is
\begin{equation}
    \frac{\delta\Delta_\beta}{|\Delta_\beta|}
    \lesssim 3.5\times10^{-11}\,.
\end{equation}
The string star radius $\tilde{r}_{\text{\tiny HP}}$ then follows from $\Delta_\beta$ via \eqref{eq:rb}; see Table~\ref{tab:HPBackgroundData}.

\subsection{Linear shooting method for the perturbations}

For fixed $D$ and $\ell$, the perturbation equations \eqref{eq:EOMPertSH} are linear in
\begin{equation}
    f(\tilde{r})=\delta\tilde{\chi}_\ell(\tilde{r}) \, ,
    \qquad
    g(\tilde{r})=\delta\tilde{\varphi}_\ell(\tilde{r}) \, ,
\end{equation}
and we exploit this rather than solving the boundary value problem \eqref{eq:EOMPertSH}, \eqref{eq:dchi_BCs_rmin}--\eqref{eq:devphi_robin} directly. We construct three initial value solutions, each integrated from $\tilde{r}_{\min}$ to the largest outer radius used in the analysis.

The first is a particular solution $(f_{\text{p}},g_{\text{p}})$ with the tidal source $\eta_\ell=\tilde{r}^{\ell}$ on, chosen to have vanishing initial data,
\begin{equation}
    f_{\text{p}}(\tilde{r}_{\min})=0 \, ,
    \qquad
    f_{\text{p}}'(\tilde{r}_{\min})=0 \, ,
    \qquad
    g_{\text{p}}(\tilde{r}_{\min})=0 \, ,
    \qquad
    g_{\text{p}}'(\tilde{r}_{\min})=0 \, .
\end{equation}
The other two are homogeneous regular solutions with the source off. One, $(f_\chi,g_\chi)$, is initialized as
\begin{equation}
    f_\chi(\tilde{r}_{\min})=\tilde{r}_{\min}^{\ell} \, ,
    \qquad
    f_\chi'(\tilde{r}_{\min})=\ell \, \tilde{r}_{\min}^{\ell-1} \, ,
    \qquad
    g_\chi(\tilde{r}_{\min})=0 \, ,
    \qquad
    g_\chi'(\tilde{r}_{\min})=0 \, ,
\end{equation}
the other, $(f_\varphi,g_\varphi)$, as
\begin{equation}
    f_\varphi(\tilde{r}_{\min})=0 \, ,
    \qquad
    f_\varphi'(\tilde{r}_{\min})=0 \, ,
    \qquad
    g_\varphi(\tilde{r}_{\min})=\tilde{r}_{\min}^{\ell} \, ,
    \qquad
    g_\varphi'(\tilde{r}_{\min})=\ell \, \tilde{r}_{\min}^{\ell-1} \, .
\end{equation}
The general regular solution with the source on, obeying \eqref{eq:dchi_BCs_rmin}--\eqref{eq:dphi_BCs_rmin}, is then
\begin{equation}
    f
    =
    f_{\text{p}}
    +
    a_\ell \, f_{\chi}
    +
    b_\ell \, f_{\varphi} \, ,
    \qquad
    g
    =
    g_{\text{p}}
    +
    a_\ell \, g_{\chi}
    +
    b_\ell \, g_{\varphi} \, ,
\end{equation}
with $a_\ell$ and $b_\ell$ fixed by the two outer conditions \eqref{eq:dchi_BC_rmax}--\eqref{eq:devphi_robin}. At an outer boundary $\tilde{r}=\tilde{r}_{\max}$, these read
\begin{equation}
    f(\tilde{r}_{\max})=0 \, ,
    \qquad
    \tilde{r}_{\max} \, g'(\tilde{r}_{\max})+(\ell+d-2) \, g(\tilde{r}_{\max})=0 \, ,
\end{equation}
and these two outer conditions form a $2\times2$ linear system  for $a_\ell$ and $b_\ell$. This is equivalent to solving the original boundary value problem, but numerically more stable.

To gauge stability we repeat the construction for three outer radii. In each of $D=4,5,6$ we use $\tilde{r}_{\max}=35,40,45$ for $\ell=2$, $\tilde{r}_{\max}=40,45,50$ for $\ell=3$, and $\tilde{r}_{\max}=45,50,55$ for $\ell=4$, always with the same inner cutoff $\tilde{r}_{\min}=10^{-3}$ as in the unperturbed problem.

\subsection{Plateau extraction and stability estimates}

For each $\tilde{r}_{\max}$, once $g(\tilde{r})=\delta\tilde{\varphi}_\ell(\tilde{r})$ is constructed, we extract $\mathcal{C}_\ell^\varphi$ (cf.~\eqref{eq:Cell_def}) from the plateau of
\begin{equation}
\label{eq:Pphi}
    P_\varphi(\tilde{r})
    =
    \tilde{r}^{\ell+d-2}g(\tilde{r}) \, ,
\end{equation}
which in the exact solution tends to $\mathcal{C}_\ell^\varphi$ at large radius. We scan windows of fixed width between the central region and the outer boundary, sampling $P_\varphi$ in each, computing the mean $\overline{P}_\varphi$, and defining the relative spread
\begin{equation}
    \epsilon_{\text{window}}
    =
    \frac{\max_i \left|P_\varphi(\tilde{r}_i)-\overline{P}_\varphi \right|}
    {\left|\overline{P}_\varphi \right|} \, .
\end{equation}
Ordering the candidate windows by this spread, we denote the five flattest by $w_n$ ($n=1,\ldots,5$, with $w_1$ flattest), write $\mathcal{C}_{\ell,n}^\varphi(\tilde{r}_{\max})$ for the $\overline{P}_\varphi$ extracted from $w_n$, and $\epsilon_n$ for its relative spread. At fixed $\tilde{r}_{\max}$ the best value is
\begin{equation}
    \mathcal{C}_{\ell,\tilde{r}_{\max}}^\varphi
    =
    \mathcal{C}_{\ell,1}^\varphi(\tilde{r}_{\max}) \, ,
\end{equation}
with plateau-variation estimate
\begin{equation}
\label{eq:Cplateau_uncertainty_fixed_rmax}
    \delta\mathcal{C}_{\tilde{r}_{\max}}^{(1)}
    =
    \max
    \left\{
    \max_{n=1,\ldots,5}
    \left|
    \mathcal{C}_{\ell,n}^\varphi(\tilde{r}_{\max})
    -
    \mathcal{C}_{\ell,1}^\varphi(\tilde{r}_{\max})
    \right| \, ,
    \quad
    \max_{n=1,\ldots,5}
    \left|
    \mathcal{C}_{\ell,n}^\varphi(\tilde{r}_{\max})
    \right|
    \epsilon_n
    \right\} \, .
\end{equation}
The first term measures the change under variation of the window among the five flattest candidates, the second the variation of $P_\varphi$ within each.

The final value of $\mathcal{C}_\ell^\varphi$ is the mean over the three outer radii,
\begin{equation}
\label{eq:Cbest}
    \mathcal{C}_{\ell,\text{best}}^\varphi
    =
    \frac{1}{3}
    \sum_{\tilde{r}_{\max}}
    \mathcal{C}_{\ell,\tilde{r}_{\max}}^\varphi \, ,
\end{equation}
and the final stability estimate is the larger of the plateau-variation estimate
\begin{equation}
    \delta\mathcal{C}^{(1)} = \max_{\tilde{r}_{\max}} \delta\mathcal{C}_{\tilde{r}_{\max}}^{(1)}
\end{equation}
and the maximum deviation from the mean \eqref{eq:Cbest},
\begin{equation}
    \delta\mathcal{C}^{(2)}
    =
    \max_{\tilde{r}_{\max}}
    \left|
    \mathcal{C}_{\ell,\tilde{r}_{\max}}^\varphi
    -
    \mathcal{C}_{\ell,\text{best}}^\varphi
    \right| \, ,
\end{equation}
that is,
\begin{equation}
\label{eq:Delta_C}
    \delta\mathcal{C}_\ell^\varphi
    =
    \max
    \left\{
    \delta\mathcal{C}^{(1)},
    \delta\mathcal{C}^{(2)}
    \right\} \, .
\end{equation}
The resulting mean value \eqref{eq:Cbest} for $\ell=2,3,4$ in $D=4,5,6$ is given in Table~\ref{tab:CphiNumerics}. Across all cases considered, the extraction is stable at the relative level
\begin{equation}
    \frac{\delta\mathcal{C}_\ell^\varphi}
    {|\mathcal{C}_\ell^\varphi|}
    \lesssim 2.5\times10^{-3}\,.
\end{equation}

\begin{table}[t!]
\centering
\small
\begin{tabular}{|c||c|c|c|}
    \hline
    $\mathcal{C}_\ell^\varphi$ & $\ell=2$ & $\ell=3$ & $\ell=4$ \\
    \hline\hline
    $D=4$
    &
    $2.20 \times 10^2$
    &
    $2.42 \times 10^3$
    &
    $5.15 \times 10^4$
    \\
    \hline
    $D=5$
    &
    $6.28 \times 10^3$
    &
    $1.68 \times 10^5$
    &
    $8.59 \times 10^6$
    \\
    \hline
    $D=6$
    &
    $4.90 \times 10^5$
    &
    $4.28 \times 10^7$
    &
    $7.28 \times 10^9$
    \\
    \hline
\end{tabular}
\caption{\small Asymptotic response coefficients
$\mathcal{C}_\ell^\varphi$ (cf.~\eqref{eq:Cell_def}) for the HP string star in $D=4,5,6$, for $\ell=2,3,4$, extracted from the plateau of \eqref{eq:Pphi}. Under variations of the plateau window and of the finite outer boundary $\tilde r_{\max}$, the maximum relative variation is approximately $2.5\times10^{-3}$. This should be regarded as a stability diagnostic rather than as a complete estimate of the numerical error; possible additional systematics are discussed below. The displayed values are
rounded to three significant figures.}
\label{tab:CphiNumerics}
\end{table}

The quantities $\delta\Delta_\beta$ and $\delta\mathcal{C}_\ell^\varphi$ quantify the stability of the asymptotic extraction under the variations described above. Specifically, $\delta\Delta_\beta$ measures the spread across the sampling window, while $\delta\mathcal{C}_\ell^\varphi$ combines the variation among the plateau windows with that obtained by varying the finite outer boundary $\tilde{r}_{\max}$ of the perturbation problem. They do not probe possible additional numerical systematics associated with the fixed inner cutoff $\tilde{r}_{\min}$, the fixed outer radius $\tilde{r}_{\max}^{\text{unpert}}$ of the unperturbed problem, the truncation of the near-origin expansions used to set the initial data, or the shooting and integration tolerances. The total numerical error may therefore be larger than these extraction-stability estimates.

Finally, since (cf.~\eqref{eq:TLN_Cell}--\eqref{eq:TLNsfinal})
\begin{equation}
\label{eq:lambda_num_best}
    \lambda_\ell
    =
    \frac{(2\ell+d-2)\Gamma(d/2)}{4\pi^{d/2}}
    \,
    \Delta_\beta^{p}
    \,
    \mathcal{C}_\ell^\varphi \, ,
    \qquad
    p=\frac{2\ell+d-2}{2} \, ,
\end{equation}
we combine the numerical stability estimates for $\mathcal{C}_\ell^\varphi$ and $\Delta_\beta$ linearly as
\begin{equation}
\label{eq:Delta_lambda_num}
    \delta\lambda_\ell
    =
    |\lambda_\ell|
    \left(
    \frac{\delta\mathcal{C}_\ell^\varphi}{|\mathcal{C}_\ell^\varphi|}
    +
    p
    \frac{\delta\Delta_\beta}{|\Delta_\beta|}
    \right) \, ,
\end{equation}
with $\Delta_\beta$ and $\mathcal{C}_\ell^\varphi$ at their best values \eqref{eq:delta_beta_best} and \eqref{eq:Cbest}. The resulting Love numbers are stable at a relative level of approximately $2.5\times10^{-3}$ or better under the variations considered here; their central values are quoted in Table~\ref{tab:HPLoveNumbers}.

\bibliography{refs}

\providecommand{\href}[2]{#2}\begingroup\raggedright\begin{thebibliography}{10}

\bibitem{Binnington:2009bb}
T.~Binnington and E.~Poisson, \emph{{Relativistic theory of tidal Love numbers}}, \href{https://doi.org/10.1103/PhysRevD.80.084018}{\emph{Phys. Rev. D} {\bfseries 80} (2009) 084018} [\href{https://arxiv.org/abs/0906.1366}{{\ttfamily 0906.1366}}].

\bibitem{Damour:2009vw}
T.~Damour and A.~Nagar, \emph{{Relativistic tidal properties of neutron stars}}, \href{https://doi.org/10.1103/PhysRevD.80.084035}{\emph{Phys. Rev. D} {\bfseries 80} (2009) 084035} [\href{https://arxiv.org/abs/0906.0096}{{\ttfamily 0906.0096}}].

\bibitem{Kol:2011vg}
B.~Kol and M.~Smolkin, \emph{{Black hole stereotyping: Induced gravito-static polarization}}, \href{https://doi.org/10.1007/JHEP02(2012)010}{\emph{JHEP} {\bfseries 02} (2012) 010} [\href{https://arxiv.org/abs/1110.3764}{{\ttfamily 1110.3764}}].

\bibitem{Hui:2020xxx}
L.~Hui, A.~Joyce, R.~Penco, L.~Santoni and A.~R. Solomon, \emph{{Static response and Love numbers of Schwarzschild black holes}}, \href{https://doi.org/10.1088/1475-7516/2021/04/052}{\emph{JCAP} {\bfseries 04} (2021) 052} [\href{https://arxiv.org/abs/2010.00593}{{\ttfamily 2010.00593}}].

\bibitem{Porto:2016zng}
R.~A. Porto, \emph{{The Tune of Love and the Nature(ness) of Spacetime}}, \href{https://doi.org/10.1002/prop.201600064}{\emph{Fortsch. Phys.} {\bfseries 64} (2016) 723} [\href{https://arxiv.org/abs/1606.08895}{{\ttfamily 1606.08895}}].

\bibitem{Charalambous:2021mea}
P.~Charalambous, S.~Dubovsky and M.~M. Ivanov, \emph{{Hidden Symmetry of Vanishing Love Numbers}}, \href{https://doi.org/10.1103/PhysRevLett.127.101101}{\emph{Phys. Rev. Lett.} {\bfseries 127} (2021) 101101} [\href{https://arxiv.org/abs/2103.01234}{{\ttfamily 2103.01234}}].

\bibitem{Charalambous:2022rre}
P.~Charalambous, S.~Dubovsky and M.~M. Ivanov, \emph{{Love symmetry}}, \href{https://doi.org/10.1007/JHEP10(2022)175}{\emph{JHEP} {\bfseries 10} (2022) 175} [\href{https://arxiv.org/abs/2209.02091}{{\ttfamily 2209.02091}}].

\bibitem{Hui:2021vcv}
L.~Hui, A.~Joyce, R.~Penco, L.~Santoni and A.~R. Solomon, \emph{{Ladder symmetries of black holes. Implications for love numbers and no-hair theorems}}, \href{https://doi.org/10.1088/1475-7516/2022/01/032}{\emph{JCAP} {\bfseries 01} (2022) 032} [\href{https://arxiv.org/abs/2105.01069}{{\ttfamily 2105.01069}}].

\bibitem{Lupsasca:2025pnt}
A.~Lupsasca, \emph{{Why there is no Love in black holes}},  \href{https://arxiv.org/abs/2506.05298}{{\ttfamily 2506.05298}}.

\bibitem{ParraMartinez:2025nvt}
J.~Parra-Martinez and A.~Podo, \emph{{Naturalness of vanishing black-hole tides}},  \href{https://arxiv.org/abs/2510.20694}{{\ttfamily 2510.20694}}.

\bibitem{Cano:2026lfl}
P.~A. Cano, M.~David, R.~Emparan and S.~Trezzi, \emph{{Love at First Loop: Quantum Deformability of Near-Extremal Black Holes}}, {\emph{to appear} (2026) }.

\bibitem{Emparan:2017qxd}
R.~Emparan, A.~Fernandez-Pique and R.~Luna, \emph{{Geometric polarization of plasmas and Love numbers of AdS black branes}}, \href{https://doi.org/10.1007/JHEP09(2017)150}{\emph{JHEP} {\bfseries 09} (2017) 150} [\href{https://arxiv.org/abs/1707.02777}{{\ttfamily 1707.02777}}].

\bibitem{Franzin:2024cah}
E.~Franzin, A.~M. Frassino and J.~V. Rocha, \emph{{Tidal Love numbers of static black holes in anti-de Sitter}}, \href{https://doi.org/10.1007/JHEP12(2024)224}{\emph{JHEP} {\bfseries 12} (2025) 224} [\href{https://arxiv.org/abs/2410.23545}{{\ttfamily 2410.23545}}].

\bibitem{Flanagan:2007ix}
E.~E. Flanagan and T.~Hinderer, \emph{{Constraining neutron star tidal Love numbers with gravitational wave detectors}}, \href{https://doi.org/10.1103/PhysRevD.77.021502}{\emph{Phys. Rev. D} {\bfseries 77} (2008) 021502} [\href{https://arxiv.org/abs/0709.1915}{{\ttfamily 0709.1915}}].

\bibitem{Cardoso:2017cfl}
V.~Cardoso, E.~Franzin, A.~Maselli, P.~Pani and G.~Raposo, \emph{{Testing strong-field gravity with tidal Love numbers}}, \href{https://doi.org/10.1103/PhysRevD.95.084014}{\emph{Phys. Rev. D} {\bfseries 95} (2017) 084014} [\href{https://arxiv.org/abs/1701.01116}{{\ttfamily 1701.01116}}].

\bibitem{Sennett:2017etc}
N.~Sennett, T.~Hinderer, J.~Steinhoff, A.~Buonanno and S.~Ossokine, \emph{Distinguishing boson stars from black holes and neutron stars from tidal interactions in inspiraling binary systems}, \href{https://doi.org/10.1103/PhysRevD.96.024002}{\emph{Phys. Rev. D} {\bfseries 96} (2017) 024002} [\href{https://arxiv.org/abs/1704.08651}{{\ttfamily 1704.08651}}].

\bibitem{Bianchi:2022qph}
M.~Bianchi and G.~Di~Russo, \emph{{2-charge circular fuzz-balls and their perturbations}}, \href{https://doi.org/10.1007/JHEP08(2023)217}{\emph{JHEP} {\bfseries 08} (2023) 217} [\href{https://arxiv.org/abs/2212.07504}{{\ttfamily 2212.07504}}].

\bibitem{Chirenti:2020bas}
C.~Chirenti, C.~Posada and V.~Guedes, \emph{Where is love? tidal deformability in the black hole compactness limit}, \href{https://doi.org/10.1088/1361-6382/abb07a}{\emph{Class. Quant. Grav.} {\bfseries 37} (2020) 195017} [\href{https://arxiv.org/abs/2005.10794}{{\ttfamily 2005.10794}}].

\bibitem{Giri:2024cks}
S.~Giri, U.~Danielsson, L.~Lehner and F.~Pretorius, \emph{{Exploring black hole mimickers: Electromagnetic and gravitational signatures of AdS black shells}}, \href{https://doi.org/10.1103/PhysRevD.111.024007}{\emph{Phys. Rev. D} {\bfseries 111} (2025) 024007} [\href{https://arxiv.org/abs/2405.08062}{{\ttfamily 2405.08062}}].

\bibitem{Rodriguez:2026iot}
M.~J. Rodr{\'\i}guez, L.~Santoni and A.~R. Solomon, \emph{{Love numbers of black holes and compact objects}},  \href{https://arxiv.org/abs/2604.08653}{{\ttfamily 2604.08653}}.

\bibitem{Chakraborty:2026qru}
S.~Chakraborty and P.~Pani, \emph{{Tidal Response of Compact Objects}},  \href{https://arxiv.org/abs/2604.08679}{{\ttfamily 2604.08679}}.

\bibitem{Susskind:1993ws}
L.~Susskind, \emph{{Some speculations about black hole entropy in string theory}},  \href{https://arxiv.org/abs/hep-th/9309145}{{\ttfamily hep-th/9309145}}.

\bibitem{Horowitz:1996nw}
G.~T. Horowitz and J.~Polchinski, \emph{{A Correspondence principle for black holes and strings}}, \href{https://doi.org/10.1103/PhysRevD.55.6189}{\emph{Phys. Rev. D} {\bfseries 55} (1997) 6189} [\href{https://arxiv.org/abs/hep-th/9612146}{{\ttfamily hep-th/9612146}}].

\bibitem{Damour:1999aw}
T.~Damour and G.~Veneziano, \emph{{Selfgravitating fundamental strings and black holes}}, \href{https://doi.org/10.1016/S0550-3213(99)00596-9}{\emph{Nucl. Phys. B} {\bfseries 568} (2000) 93} [\href{https://arxiv.org/abs/hep-th/9907030}{{\ttfamily hep-th/9907030}}].

\bibitem{Chen:2021dsw}
Y.~Chen, J.~Maldacena and E.~Witten, \emph{{On the black hole/string transition}}, \href{https://doi.org/10.1007/JHEP01(2023)103}{\emph{JHEP} {\bfseries 01} (2023) 103} [\href{https://arxiv.org/abs/2109.08563}{{\ttfamily 2109.08563}}].

\bibitem{Ceplak:2023afb}
N.~{\v{C}}eplak, R.~Emparan, A.~Puhm and M.~Toma{\v{s}}evi{\'c}, \emph{{The correspondence between rotating black holes and fundamental strings}}, \href{https://doi.org/10.1007/JHEP11(2023)226}{\emph{JHEP} {\bfseries 11} (2023) 226} [\href{https://arxiv.org/abs/2307.03573}{{\ttfamily 2307.03573}}].

\bibitem{Silva:2017uqg}
H.~O. Silva, J.~Sakstein, L.~Gualtieri, T.~P. Sotiriou and E.~Berti, \emph{{Spontaneous scalarization of black holes and compact stars from a Gauss-Bonnet coupling}}, \href{https://doi.org/10.1103/PhysRevLett.120.131104}{\emph{Phys. Rev. Lett.} {\bfseries 120} (2018) 131104} [\href{https://arxiv.org/abs/1711.02080}{{\ttfamily 1711.02080}}].

\bibitem{Horowitz:1997jc}
G.~T. Horowitz and J.~Polchinski, \emph{{Selfgravitating fundamental strings}}, \href{https://doi.org/10.1103/PhysRevD.57.2557}{\emph{Phys. Rev. D} {\bfseries 57} (1998) 2557} [\href{https://arxiv.org/abs/hep-th/9707170}{{\ttfamily hep-th/9707170}}].

\bibitem{Charalambous:2024tdj}
P.~Charalambous, \emph{{Love numbers and Love symmetries for p-form and gravitational perturbations of higher-dimensional spherically symmetric black holes}}, \href{https://doi.org/10.1007/JHEP04(2024)122}{\emph{JHEP} {\bfseries 04} (2024) 122} [\href{https://arxiv.org/abs/2402.07574}{{\ttfamily 2402.07574}}].

\bibitem{Katagiri:2024fpn}
T.~Katagiri, V.~Cardoso, T.~Ikeda and K.~Yagi, \emph{{Tidal response beyond vacuum general relativity with a canonical definition}}, \href{https://doi.org/10.1103/PhysRevD.111.084081}{\emph{Phys. Rev. D} {\bfseries 111} (2025) 084081} [\href{https://arxiv.org/abs/2410.02531}{{\ttfamily 2410.02531}}].

\bibitem{Cano:2025zyk}
P.~A. Cano, \emph{{Love numbers beyond GR from the modified Teukolsky equation}}, \href{https://doi.org/10.1007/JHEP07(2025)152}{\emph{JHEP} {\bfseries 07} (2025) 152} [\href{https://arxiv.org/abs/2502.20185}{{\ttfamily 2502.20185}}].

\bibitem{Emparan:2024mbp}
R.~Emparan, M.~Sanchez-Garitaonandia and M.~Toma{\v{s}}evi{\'c}, \emph{{String theory in a pinch: resolving the Gregory-Laflamme singularity}}, \href{https://doi.org/10.1007/JHEP02(2025)104}{\emph{JHEP} {\bfseries 02} (2025) 104} [\href{https://arxiv.org/abs/2411.14998}{{\ttfamily 2411.14998}}].

\bibitem{Atick:1988si}
J.~J. Atick and E.~Witten, \emph{{The Hagedorn Transition and the Number of Degrees of Freedom of String Theory}}, \href{https://doi.org/10.1016/0550-3213(88)90151-4}{\emph{Nucl. Phys. B} {\bfseries 310} (1988) 291}.

\bibitem{Balthazar:2022hno}
B.~Balthazar, J.~Chu and D.~Kutasov, \emph{{On small black holes in string theory}}, \href{https://doi.org/10.1007/JHEP03(2024)116}{\emph{JHEP} {\bfseries 03} (2024) 116} [\href{https://arxiv.org/abs/2210.12033}{{\ttfamily 2210.12033}}].

\bibitem{Bedroya:2024igb}
A.~Bedroya and D.~H. Wu, \emph{{String stars in d {\ensuremath{\geq}} 7}}, \href{https://doi.org/10.1007/JHEP09(2025)204}{\emph{JHEP} {\bfseries 09} (2025) 204} [\href{https://arxiv.org/abs/2412.19888}{{\ttfamily 2412.19888}}].

\bibitem{Poisson_2014}
E.~Poisson and C.~Will, \emph{{Gravity: Newtonian, Post-Newtonian, Relativistic}}. Cambridge Univ. Pr., 2014.

\bibitem{Kodama:2003jz}
H.~Kodama and A.~Ishibashi, \emph{{A Master equation for gravitational perturbations of maximally symmetric black holes in higher dimensions}}, \href{https://doi.org/10.1143/PTP.110.701}{\emph{Prog. Theor. Phys.} {\bfseries 110} (2003) 701} [\href{https://arxiv.org/abs/hep-th/0305147}{{\ttfamily hep-th/0305147}}].

\bibitem{Ishibashi:2011ws}
A.~Ishibashi and H.~Kodama, \emph{{Perturbations and Stability of Static Black Holes in Higher Dimensions}}, \href{https://doi.org/10.1143/PTPS.189.165}{\emph{Prog. Theor. Phys. Suppl.} {\bfseries 189} (2011) 165} [\href{https://arxiv.org/abs/1103.6148}{{\ttfamily 1103.6148}}].

\bibitem{Barbosa:2025uau}
S.~Barbosa, P.~Brax, S.~Fichet and L.~de~Souza, \emph{{Running Love numbers and the Effective Field Theory of gravity}}, \href{https://doi.org/10.1088/1475-7516/2025/07/071}{\emph{JCAP} {\bfseries 07} (2025) 071} [\href{https://arxiv.org/abs/2501.18684}{{\ttfamily 2501.18684}}].

\bibitem{Myers:1987qx}
R.~C. Myers, \emph{{Superstring Gravity and Black Holes}}, \href{https://doi.org/10.1016/0550-3213(87)90402-0}{\emph{Nucl. Phys. B} {\bfseries 289} (1987) 701}.

\bibitem{Chen:2021qrz}
Y.~Chen, \emph{{Revisiting $R^4$ higher curvature corrections to black holes}},  \href{https://arxiv.org/abs/2107.01533}{{\ttfamily 2107.01533}}.

\bibitem{Callan:1988hs}
C.~G. Callan, Jr., R.~C. Myers and M.~J. Perry, \emph{{Black Holes in String Theory}}, \href{https://doi.org/10.1016/0550-3213(89)90172-7}{\emph{Nucl. Phys. B} {\bfseries 311} (1989) 673}.

\bibitem{Moura:2006pz}
F.~Moura and R.~Schiappa, \emph{{Higher-derivative corrected black holes: Perturbative stability and absorption cross-section in heterotic string theory}}, \href{https://doi.org/10.1088/0264-9381/24/2/006}{\emph{Class. Quant. Grav.} {\bfseries 24} (2007) 361} [\href{https://arxiv.org/abs/hep-th/0605001}{{\ttfamily hep-th/0605001}}].

\bibitem{Cano:2021rey}
P.~A. Cano and A.~Ruip{\'e}rez, \emph{{String gravity in D=4}}, \href{https://doi.org/10.1103/PhysRevD.105.044022}{\emph{Phys. Rev. D} {\bfseries 105} (2022) 044022} [\href{https://arxiv.org/abs/2111.04750}{{\ttfamily 2111.04750}}].

\bibitem{Cano:2022wwo}
P.~A. Cano, B.~Ganchev, D.~R. Mayerson and A.~Ruip{\'e}rez, \emph{{Black hole multipoles in higher-derivative gravity}}, \href{https://doi.org/10.1007/JHEP12(2022)120}{\emph{JHEP} {\bfseries 12} (2022) 120} [\href{https://arxiv.org/abs/2208.01044}{{\ttfamily 2208.01044}}].

\bibitem{Wang:2026qst}
L.~Wang, L.~Lehner, M.~Micol and R.~Sturani, \emph{{Matching Tidal Deformability (Wilson) Coefficients to Black Hole Love Numbers in Higher-Curvature Gravity}},  \href{https://arxiv.org/abs/2604.04259}{{\ttfamily 2604.04259}}.

\bibitem{Cano:2026hlv}
P.~A. Cano, F.~Fucito, J.~F. Morales and A.~Ruip{\'e}rez, \emph{{Gravitational waveforms from binaries in higher-derivative gravity: a Love story}},  \href{https://arxiv.org/abs/2606.07070}{{\ttfamily 2606.07070}}.

\bibitem{YipLeung2017}
K.~L.~S. Yip and P.~T. Leung, \emph{{Tidal Love numbers and moment--Love relations of polytropic stars}}, \href{https://doi.org/10.1093/mnras/stx2363}{\emph{Mon. Not. Roy. Astron. Soc.} {\bfseries 472} (2017) 4965} [\href{https://arxiv.org/abs/1709.02469}{{\ttfamily 1709.02469}}].

\bibitem{Bena:2022rna}
I.~Bena, E.~J. Martinec, S.~D. Mathur and N.~P. Warner, \emph{{Fuzzballs and Microstate Geometries: Black-Hole Structure in String Theory}},  \href{https://arxiv.org/abs/2204.13113}{{\ttfamily 2204.13113}}.

\bibitem{Bianchi:2023sfs}
M.~Bianchi, G.~Di~Russo, A.~Grillo, J.~F. Morales and G.~Sudano, \emph{{On the stability and deformability of top stars}}, \href{https://doi.org/10.1007/JHEP12(2023)121}{\emph{JHEP} {\bfseries 12} (2023) 121} [\href{https://arxiv.org/abs/2305.15105}{{\ttfamily 2305.15105}}].

\bibitem{DiRusso:2024hmd}
G.~Di~Russo, F.~Fucito and J.~F. Morales, \emph{{Tidal resonances for fuzzballs}}, \href{https://doi.org/10.1007/JHEP04(2024)149}{\emph{JHEP} {\bfseries 04} (2024) 149} [\href{https://arxiv.org/abs/2402.06621}{{\ttfamily 2402.06621}}].

\bibitem{Ceplak:2024dxm}
N.~{\v{C}}eplak, R.~Emparan, A.~Puhm and M.~Toma{\v{s}}evi{\'c}, \emph{{Size and shape of rotating strings and the correspondence to black holes}}, \href{https://doi.org/10.1007/JHEP06(2025)099}{\emph{JHEP} {\bfseries 06} (2025) 099} [\href{https://arxiv.org/abs/2411.18690}{{\ttfamily 2411.18690}}].

\bibitem{Santos:2024ycg}
J.~E. Santos and Y.~Zigdon, \emph{{Self gravitating spinning string condensates}}, \href{https://doi.org/10.1007/JHEP07(2024)217}{\emph{JHEP} {\bfseries 07} (2024) 217} [\href{https://arxiv.org/abs/2403.20332}{{\ttfamily 2403.20332}}].

\bibitem{Seitz:2025wpc}
J.~Seitz and E.~Y. Urbach, \emph{{A spin on Hagedorn temperatures and string stars}}, \href{https://doi.org/10.1007/JHEP06(2026)245}{\emph{JHEP} {\bfseries 06} (2026) 245} [\href{https://arxiv.org/abs/2510.17951}{{\ttfamily 2510.17951}}].

\end{thebibliography}\endgroup

\end{document}